\documentclass[12pt]{article}
\usepackage[cmtip,arrow]{xy}\usepackage{pb-diagram,pb-xy}%

\input xy
\xyoption{all}
\input xy
\xyoption{all}
\usepackage{heck}
\usepackage{cite}
\usepackage{graphicx}
\usepackage{makeidx}
\usepackage{multicol}
\usepackage{amsfonts}
\usepackage{mathrsfs}
\usepackage{amssymb}
\usepackage{amsmath}%

\providecommand{\U}[1]{\protect\rule{.1in}{.1in}}
\numberwithin{equation}{section}

\newcommand{\bea}{\begin{eqnarray}}
\newcommand{\eea}{\end{eqnarray}}
\newcommand{\be}{\begin{equation}}
\newcommand{\ee}{\end{equation}}

\newcommand{\bem}{\begin{pmatrix}}
\newcommand{\eem}{\end{pmatrix}}

\def\a{\alpha}

\def\d{\delta}

\def\U{\Upsilon}

\def\cc{{\cal C}}

\def\ch{{\cal H}}

\def\cn{{\cal N}}

\def\cp{{\cal P}}

\def\ct{{\cal T}}

\def\cw{{\cal W}}

\def \Z {{\mathbb Z}}
\def \C {{\mathbb C}}

\def\fd{\mathfrak{d}}

\xyoption{arc}

\date{July, 2012}

\institution{SISSA}{\centerline{  Scuola Internazionale Superiore di Studi Avanzati, via Bonomea 265,I-34100 Trieste, ITALY}}

\title{Half--Hypers and Quivers}
\authors{Sergio Cecotti\worksat{\SISSA}\footnote{e-mail: {\tt cecotti@sissa.it}} and Michele Del Zotto \worksat{\SISSA}\footnote{e-mail: {\tt eledelz@gmail.com}}%
}

\abstract{We study systematically the BPS spectra of $\cn=2$ SYM coupled to \emph{half}--hypermultiplets, the basic example being $E_7$ SYM coupled to a half--hyper in the $\mathbf{56}$ irrepr.  In order to do this, we determine the BPS quivers with superpotential of such $\cn=2$ models using a new technique we introduce. 
The computation of the BPS spectra in the various chambers is then reduced to the Representation Theory of the resulting quivers. We use the quiver description to study the BPS spectrum at both strong and weak coupling. 
The following models are discussed in detail:  $SU(6)$ SYM coupled to a $\tfrac{1}{2}\,\mathbf{20}$, $SO(12)$ SYM coupled to a $\tfrac{1}{2}\,\mathbf{32}$, and $E_7$ SYM coupled to a $\tfrac{1}{2}\,\mathbf{56}$. For models with gauge group $SU(2)\times SO(2n)$ and matter in the $\tfrac{1}{2}(\mathbf{2},\mathbf{2n})$ we find  strongly coupled chambers with a BPS spectrum consisting of just finitely many hypermultiplets.
}

\begin{document}

\maketitle

\tableofcontents
\newpage

\section{Introduction}\label{intt}

The non--pertubative physics of \emph{non}--vector--like 4d gauge theories is notoriously a subtle and tricky subject \cite{wittenZ2}. Their peculiar features may be ultimately traced to the fact that such theories do not have gauge--invariant mass terms for all fermions. 

In the context of $\mathcal{N}=2$ supersymmetric gauge theories, the corresponding problem   is to understand the physics of $\cn=2$ SYM coupled to an \textit{odd} number of \emph{half}--hypermultiplets, where by a `half'--hypermultiplet we mean a hypermultiplet in an irreducible \emph{quaternionic} representation of the gauge group $G$. Although many classical theories of this class become inconsistent at the quantum level \cite{wittenZ2}, some models of this form are known to be fully consistent QFTs since they may be engineered in Type IIB superstring (see \cite{Tack} for a recent discussion). From the field theory side such theories appear to live on the verge of inconsistency, and own their existence to peculiar mechanisms which look rather miraculous. 

Luckily, in the $\cn=2$ supersymmetric case we have at our disposal many powerful techniques to address non--perturbative questions \cite{SW1,SW2,Gaiotto,GMN09,GMN10,GMN12}. An especially simple and elegant method is the quiver approach \cite{CV11,ACCERV1,ACCERV2} (for previous work see \cite{Denef00,DM07,DM,Dia99,DFR1,DFR2,FM00,Fiol,Denef,FHHI02,FHH00,Feng:2001xr,HK05,FHKVW05,Feng:2005gw} and the nice review \cite{review}) which, for the convenience of the reader, we briefly review in section 2 below. The quiver method applies to a large class of $\cn=2$ gauge theories (called \textit{`quiver gauge theories'} \cite{CV11}), and consists in mapping the computation of the non--perturbative BPS spectrum into the Representation Theory of the quiver(s) associated with the given gauge theory.

The preliminary step in this approach is to determine the quiver class associated to the $\cn=2$ quiver theory of interest. For 4d $\cn=2$ theories which have a Type IIB engineering, the quiver  class is determined, in principle, by the $2d/4d$ correspondence \cite{CNV}. To determine the quiver  class directly from the geometry is quite difficult in general. In refs.\!\cite{CV11,ACCERV2} a simple prescription was given to construct the quiver for $\cn=2$ SYM (with a simply--laced gauge group $G$) coupled to \emph{full} hypermultiplets in arbitrary representations of $G$. Ignoring some subtleties with large representations of $G$ (see \cite{cattoy}), their strategy is very simple \cite{CNV,ACCERV2}: one takes the mass--parameters of all hypermultiplets to be very large, and exploits the fact that in the decoupling limit we must get back the known quiver for pure $G$ SYM. This strategy cannot be used in presence of half--hypermultiplets since there is no mass parameter to take to infinity. The determination of the quiver class for $\cn=2$ SYM coupled to half--hypers was left open in ref.\!\cite{ACCERV2}.

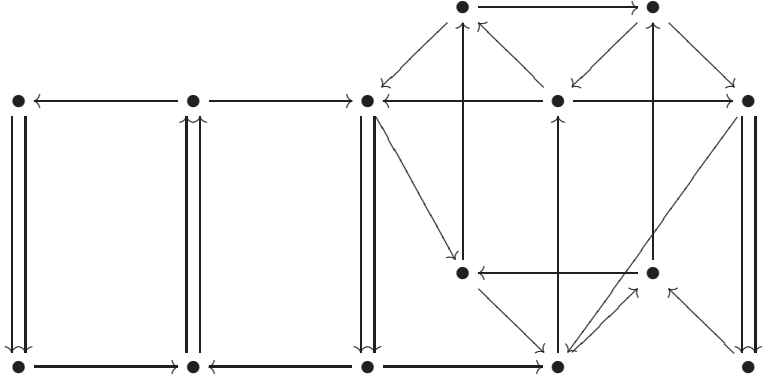
\begin{figure}
\begin{equation*}\begin{gathered}
\xymatrix{
&&&&&\bullet \ar[ld] \ar[rr] && \bullet \ar[dl]\ar[dr]&\\
\bullet \ar@<0.5ex>@{->}[ddd]\ar@<-0.5ex>@{->}[ddd]&& \bullet \ar[ll] \ar[rr]&&\bullet  \ar@<0.5ex>@{->}[ddd]\ar@<-0.5ex>@{->}[ddd] \ar[ddr]& & \bullet \ar[lu]\ar[ll] \ar[rr]&&\bullet \ar@<0.5ex>@{->}[ddd]\ar@<-0.5ex>@{->}[ddd]\ar[dddll]\\
&&&&& &  &&\\
&& &&&\bullet \ar[uuu]\ar[dr] & & \bullet \ar[ll]\ar[uuu]&\\
\bullet \ar[rr]&&\bullet \ar@<0.5ex>@{->}[uuu]\ar@<-0.5ex>@{->}[uuu]&&\bullet \ar[ll] \ar[rr]& & \bullet \ar[ur]\ar[uuu] &&\bullet \ar[ul]\\
}
\end{gathered}\end{equation*}
\caption{The quiver of $E_7$ SYM coupled to half a $\textbf{56}$.}\label{eeee777}\end{figure}

In this paper we shall fill the gap, and describe the quiver class of the $\cn=2$ gauge theories with half--hypermultiplets. 

Of course, the issue of the quiver class makes sense only for \emph{quiver theories}. Before embarking in the quest for the quiver, it is wise to ask whether a quiver \emph{exists} at all. The question is not frivolous: the \emph{non}--quiver $\cn=2$ theories are believed to have no mass deformation, and the models we consider share this `dangerous' property, so are potentially quiver--less. Luckily, the most interesting examples of gauge theories with half--multiplets have a Type IIB engineering, and for these theories the existence of a quiver is garanteed by the $2d/4d$ correspondence \cite{CNV}. This happens, for instance, for $E_7$ SYM coupled to $\tfrac{1}{2}\,\mathbf{56}$ whose Type IIB geometry is given by the elegant hypersurface in $\mathbb{C}^*\times \mathbb{C}^3$ \cite{Tack}
\begin{equation}\label{hypere7}
  z\,x_2+\frac{1}{z}=W_{E_7}(x_1,x_2,x_3;w_k)
\end{equation}
where 
\begin{equation}
 W_{E_7}(x_1,x_2,x_3;w_k)=x_1^3x_2+x_2^3+x_3^2+\text{lower order}
\end{equation}
is the versal deformation of the $E_7$ canonical singularity. The simple form of this geometry reflects the fact that the $\mathbf{56}$ is the \emph{fundamental} representation for $E_7$; in facts, this is the only quantum consistent model with a half--hyper in the  fundamental representation. The quiver $Q$ and its superpotential $\cw$ for this theory are defined in terms of the Picard--Lefshetz geometry \cite{CV92} of the hypersurface \eqref{hypere7} \cite{CNV,cattoy}; however, the explicit computation of the pair $(Q,\cw)$ from the Type IIB geometry is quite involved, unless only the germ at the origin of the hypersurface  matters (as was the case for the models studied in \cite{CNV,arnold,arnold2,arnold3}).  
 Similar arguments may be used to prove \emph{a priori} the existence of a quiver for other asymptotically--free half--hyper gauge theories; the quivers may then be explicitly constructed along the lines of the present paper. For instance, figure \ref{eeee777} represents a quiver for the $E_7$ theory coupled to half a fundamental.

\medskip

Having determined the quiver (with superpotential $\cw$) we may use it to get highly non--trivial informations about the non--perturbative physics of the model, especially at very strong coupling.  In particular, we show that in the  models with a gauge group $G=SU(2)\times SO(2n)$ coupled to one half--hyper in the $(\mathbf{2},\mathbf{2n})$ there are strongly coupled chambers in which the BPS spectrum consists only of finitely many hypermultiplets; we expect this property to hold also for other models we consider, but we leave that question open. 
\medskip

The paper is organized as follows. In section 2, after a quick review of the quiver approach to fix the notations, we introduce the Dirac integrality conditions (\S.\,\ref{Dirrrac}), describe our strategy to determine the quivers (\S.\,\ref{strategy}), and give a more precise description of the class of theories of interest. In section 3 we consider half--hypermultiplets in $SU(2)^k$ gauge theories, along the lines of \cite{CV11} and \cite{ACCERV1}. In section 4 we construct the pairs $(Q,\cw)$ for the various theories of interest: first the models with gauge groups $G=SU(2)\times SO(2n)$ coupled to $\tfrac{1}{2}\,(\mathbf{2},\mathbf{2n})$ hypers --- which are the basic building blocks for all subsequent constructions --- and then the models with gauge group $G$ simple (and simply--laced). In section 5 we discuss the strong coupling regime of such theories. Technicalities are confined in four Appendices.

\section{Quivers, BPS states, and all that}  

\subsection{Quivers and representations}
For a \textit{quiver} 4d $\cn=2$ theory (see refs.\!\cite{CV11,ACCERV1,ACCERV2,Denef,cattoy} for details) the BPS particles correspond to the \textsc{susy} vacua of a quiver 1d SQM which is specified by a quiver $Q$ and a superpotential $\cw$. 

Let $\Gamma\equiv\bigoplus_{i=1}^m \mathbb{Z}\,e_i$ be the lattice of the (quantized) conserved charges of the 4d theory. To each generator $e_i$ of $\Gamma$ there corresponds a node of $Q$, and the node $e_i$ is connected to node $e_j$ by $\langle e_i, e_j\rangle_\text{Dirac}$ arrows, where $\langle e_i,e_j\rangle_\text{Dirac}$ is the Dirac electromagnetic symplectic pairing between the charges $e_i$ and $e_j$ (a negative number of arrows means arrows in the opposite direction). In particular, flavor charges correspond to null--eigenvectors of the exchange matrix of the quiver, $$B_{ij}\equiv \langle e_i, e_j\rangle_\text{Dirac}.$$ 

Let $\gamma=\sum_i N_i\, e_i\in \Gamma$ ($N_i\geq 0$) be the charge vector of a BPS state. It corresponds to a \textsc{susy} vacuum\footnote{\ More precisely, it corresponds to the quantization of a continuous family of such vacua. If the vacuum is not rigid, the quantization of the corresponding moduli produces the 4d spin of the BPS particle \cite{ACCERV2}.} of the 1d theory with gauge group $U(N_i)$ at the $i$--th node of $Q$ and with a chiral Higgs (super)field in the bifundamental representation $(\boldsymbol{\overline{N}_i},\boldsymbol{N_j})$ for each arrow $i\rightarrow j$. The Higgs fields are subjected to the \textsc{susy} relations $\partial\cw=0$, where the superpotential $\cw$, by gauge invariance, must be a linear combination of the closed oriented cycles on $Q$.
A classical vacuum of the 1d theory is then nothing else than a representation of $Q$ satisfying the Jacobian relations $\partial\cw=0$ or, equivalently, a module of the quotient algebra $\mathbb{C}Q/(\partial\cw)$. We write $\mathsf{rep}(Q,\cw)$ for the category of such representations. The 4d charge vector $\gamma$ is just the dimension vector $$\dim X=\sum_i \dim X_i\:e_i\in \Gamma$$ of the corresponding representation $X\in\mathsf{rep}(Q,\cw)$.\smallskip

The pair $(Q,\cw)$ is not unique; indeed two pairs $(Q,\cw)$ and $(Q^\prime,\cw^\prime)$ related by SQM Seiberg duality \cite{Seiberg}\!\cite{FHHI02,FHH00} are physically equivalent. The positive lattice generators $e_i$ of the quiver $Q$ are linear combinations of the positive generators $e^\prime_i$ of $Q^\prime$. The Seiberg equivalence $$(Q,\cw)\sim_S (Q^\prime,\cw^\prime)$$ exactly corresponds to the mutation--equivalence of  quivers with superpotential in the sense of Derksen--Weyman--Zelevinsky \cite{derksen1}. Then one speaks of the \textit{mutation--class} of quivers associated to a given 4d $\cn=2$ theory; except for theories with gauge group $SU(2)^k$, the mutation--class is infinite \cite{CV11}.

The central charge of the 4d $\cn=2$ \textsc{susy} algebra is a conserved charge, and hence may be expanded in the complete basis $\{e_i\}$. This gives a linear map $Z(\cdot)\colon\Gamma\rightarrow \C$, and a particle with quantum numbers $\gamma=\sum_i N_i\, e_i\in \Gamma$ will have a central charge 
\begin{equation}Z(\gamma)=\sum_i N_i\, Z_i\qquad \text{where } Z_i\equiv Z(e_i),\end{equation}
 and a mass $M\geq |Z(\gamma)|$, with equality iff it is a BPS state. $Z(\cdot)$ depends on all the parameters of the theory (couplings, masses, Coulomb branch parameters, \textit{etc.}), and hence encodes the physical regime in which we study the theory\footnote{\ $Z(\cdot)$ also specifies the couplings of the 1d SQM model corresponding to the given charge vector $\gamma$ \cite{Denef}.}. Depending on the particular regime, we have to use the appropriate quiver(s) in the mutation--class. In a regime specified by the central charge $Z(\cdot)$, the positive generators $e_i$ of an admissible quiver  must satisfy the condition
\begin{equation}
 0\leq \arg Z(e_i)<\pi\qquad\text{for all }i.
\end{equation}
Under this condition, we extend the central charge to  
$X\in\mathsf{rep}(Q,\cw)$ by setting 
\begin{equation}Z(X)\equiv Z(\dim X)=\sum_i (\dim X)_i\, Z_i,
\end{equation}
 so that, for all non--zero $X$, we have $Z(X)\neq 0$ and $0\leq \arg Z(X)<\pi$. A representation $X\in\mathsf{rep}(Q,\cw)$ is said to be \textsc{stable} iff for all non--zero proper subrepresentation $Y$
\begin{equation}
 0\leq \arg Z(Y) < \arg Z(X). 
\end{equation}

One shows that a representation $X$ corresponds to a \textsc{susy} vacuum of the 1d theory specified by $Z(\cdot)$ if and only if it is stable \cite{Denef}\!\cite{CV11,ACCERV1,ACCERV2}. A stable representation is, in particular, a \textit{brick}  \cite{king,keller}, that is, 
\begin{equation}X\ \text{stable}\quad \Rightarrow\quad\mathrm{End}\,X=\C.\end{equation} 
A $k$--dimensional family of stable representations corresponds to a $\cn=2$ supermultiplet of maximal spin $(k+1)/2$ \cite{ACCERV2}; so a rigid stable representation corresponds to a BPS hypermultiplet, while a one--parameter family to a vector multiplet.

\subsection{Electric and magnetic weights}\label{magneticweights}

\subsubsection{Dirac integrality conditions}\label{Dirrrac}

Consider a quiver $\cn=2$ gauge theory with gauge group $G$ (which we take to be simply--laced of rank $r$). We pick a particular pair $(Q,\cw)$ in the corresponding Seiberg mutation--class which is appropriate for the weak coupling regime of its Coulomb branch.  $\mathsf{rep}(Q,\cw)$ should contain, in particular, one--parameter families of representations corresponding to the massive $W$--boson vector--multiplets which are in one--to--one correspondence with the positive roots of $G$. We write $\delta_a$ ($a=1,2,\dots, r$) for the charge (\textit{i.e.}\! dimension) vector of the $W$--boson associated to the \emph{simple--root} $\alpha_a$ of $G$.

At a generic point in the Coulomb branch we have an unbroken $U(1)^r$ symmetry.
The $U(1)^r$ electric charges, properly normalized so that they are integral for all states, are given by the fundamental coroots\footnote{\ Here and below $\mathfrak{h}$ stands for the Cartan subalgebra of the complexified  Lie algebra of the gauge group $G$.} $\alpha_a^\vee\in\mathfrak{h}$ ($a=1,2,\dots,r$). The $a$--th electric charge of the $W$--boson associated to $b$--th simple root $\alpha_b$ then is
\begin{equation}
q_a=\alpha_b(\alpha^\vee_a)=C_{ab}, \qquad \text{(the Cartan matrix of }G).
\end{equation}
Therefore the vector in $\Gamma\otimes \mathbb{Q}$ corresponding to the $a$--th unit electric charge is
\begin{equation}\mathfrak{q_a}=(C^{-1})_{ab}\,\delta_b.
\end{equation}
Then the magnetic weights (charges) of a representation $X$ are given by
\begin{equation}\label{magneticxxx}
m_a(X)\equiv \langle \dim X, \mathfrak{q}_a\rangle_\text{Dirac}=(C^{-1})_{ab}\,B_{ij}\,(\dim X)_i\,(\delta_b)_j.
\end{equation} 

Dirac quantization requires the $r$ linear forms $m_a(\cdot)$ to be \emph{integral} \cite{cattoy}. This integrality condition is quite a strong constraint on the quiver $Q$, and will be our main criterion to determine it.\medskip

\subsubsection{The weak coupling `perturbative' (sub)category}
At weak coupling, $g_\mathrm{YM}\rightarrow 0$, the central charge takes the classical form \cite{cattoy}
\begin{equation}
 Z(X)= -\frac{1}{g^2_\mathrm{YM}}\,\sum_i C_a\,m_a(X)+O(1)\qquad C_a>0.
\end{equation}
It is convenient to define the light category, $\mathscr{L}(Q,\cw)$, as the subcategory of the representations $X\in\mathsf{rep}(Q,\cw)$ with
$m_a(X)=0$ for all $a$ such that all their subrepresentations have $m_a(Y)\leq 0$. All BPS states with bounded mass in the weak coupling limit $g_\mathrm{YM}\rightarrow 0$ correspond to representations in $\mathscr{L}(Q,\cw)$, and, in facts, for a $\cn=2$ theory which has a weakly coupled Lagrangian description the stable objects of $\mathscr{L}(Q,\cw)$ precisely correspond to the perturbative states. They are just the gauge bosons, making precisely one copy of the adjoint of $G$, together with finitely many hypermultiplets transforming in definite representations of $G$. The detailed structure of $\mathscr{L}(Q,\cw)$ is described in \cite{cattoy}.

\subsection{Decoupling strategies to determine $Q$ and $\cw$}
The limit $g_\mathrm{YM}\rightarrow 0$ is a particular decoupling limit. In general \cite{cattoy} a decoupling limit is specified by a set of linear functions $$\lambda_A\colon \Gamma \rightarrow \mathbb{Z},\qquad (A=1,2,\dots, s)$$ and the subcategory $\mathscr{A}\subset \mathsf{rep}(Q,\cw)$ of representations which \emph{do not} decouple (that is, which have bounded masses in the decoupling limit) is given by the representations $X\in\mathsf{rep}(Q,\cw)$ with $\lambda_A(X)=0$ for all $A$, while for all their subrepresentations $Y$ one has $\lambda_A(Y)\leq 0$ \cite{cattoy}.

An especially simple case is when one has
$\lambda_A(e_i)\geq 0$ for all positive generators $e_i$ of $Q$ and all $A=1,2,\dots, s$. In this positive case, $$\mathscr{A}=\mathsf{rep}(Q_0,\cw\big|_{Q_0}),$$ where $Q_0$ is the full subquiver of $Q$ over the nodes $e_i$ such that $\lambda_A(e_i)=0$ for all $A$.

\subsubsection{The `massive quark' procedure}
Consider a quark of mass $m$ in the representation $R$ of the gauge group $G$ coupled to some  $\cn=2$ system. Since the quark is a hypermultiplet, one may always find a quiver $Q$ in the mutation class such that a given state of the quark (with known electric weights) corresponds to a simple representation $S_{i_0}\in \mathsf{rep}(Q,\cw)$ whose dimension vector is the positive generator $e_{i_0}$ of $Q$. For such a quiver, the decoupling limit $m\rightarrow\infty$ corresponds to the single positive $\lambda(\cdot)$ with
\begin{equation}
\lambda(e_i)= \begin{cases}1 &i=i_0\\
0 & i\neq i_0.\end{cases}
\end{equation}
In this case the full subquiver $Q_0$ is obtained from $Q$ just by erasing the $i_0$--th node. $Q_0$ must belong to the mutation class of the QFT which residues after decoupling the massive quark. For a Lagrangian $\cn=2$ theory with only \emph{full} quarks, we may repeat this decoupling procedure until we remains with pure $G$  SYM, whose quiver--class we know (for $G$ simply--laced). We can invert the decoupling procedure as follows \cite{CV11, ACCERV2}. If we know the subquiver $Q_0$ and the electric/magnetic charge of its nodes $e_i$,  to reconstruct $Q$ we add an extra node $e_{i_0}$, which represents the given state of the massive quark, whose electric weights we also know. 
Then, using the known electric/magnetic charges of all nodes $e_i$, $e_{i_o}$ we compute the Dirac pairing $\langle e_{i_0}, e_i\rangle_\text{Dirac}$,
and we connect the new node to the $i$--th node of $Q_0$ by $\langle e_{i_0}, e_i\rangle_\text{Dirac}$ arrows. The resulting quiver is our $Q$. This is the `massive quark' procedure of refs.\!\cite{CV11, ACCERV2} which gives the correct quiver provided the $\cn=2$ theory with gauge group $G$ and the given quark representation $R$ is a \emph{quiver} gauge theory which, in particular, requires $R$ to be small enough so that the coupled theory remains asymptotically free (see \cite{cattoy}).

\subsubsection{The `Higgs' procedure (our strategy)}\label{strategy}
The `massive quark'  procedure is not available in presence of half--hypers, since there is no mass parameter $m$ to send to infinity. However we may still proceed as follow. After having decoupled all \emph{full} quarks using the previous procedure, we remain with SYM coupled to half--hypers only. This theory has no flavor symmetry, so that $\det B\neq 0$, and the number of nodes of $Q$ is $2\,r$, equal to the number of electric and magnetic charges of $G$.

Suppose we know the dimension vectors $\delta_a$ of the simple--root $W$--bosons; then, using eqn.\eqref{magneticxxx}, we can determine the electric/magnetic charges of each node of $Q$; the charges are all linearly independent since $\det B\neq 0$. Now let us switch on a large v.e.v.\! of the complex adjoint scalar in the vector multiplet,
$\langle \Phi\rangle\in\mathfrak{h}$, of the form 
\begin{equation}\label{simplecase}\langle \Phi\rangle=i \,t \,\varphi\qquad \text{where }\ \alpha_a(\varphi)= \begin{cases}1 & a=a_0,\\
0 & a\neq a_0,\end{cases}
\end{equation}
 and take the limit $t\rightarrow\infty$. The $a_0$--th simple--root $W$--boson becomes infinitely massive and it decouples together with all states having an electric weight $\varrho$ such that $\varrho(\varphi)\neq 0$. Up to IR free photons, we remain with an $\cn=2$ theory with gauge group $G^\prime\subset G$ having the Dynkin diagram obtained by erasing the $a_0$--th node from the Dynkin graph of $G$; in general the residual theory will also contain half--hypers in suitable representations $R^\prime$ of $G^\prime$.

The Higgs decoupling limit \eqref{simplecase} corresponds to taking $\lambda(\cdot)$ equal to the following combinations of the electric charges $q_{a_0}(\cdot)$
\begin{equation}
 \lambda(X)= (C^{-1})_{a_0 a}\,q_a(X).
\end{equation}
 If $\lambda(e_i)\geq 0$ for all nodes of $Q$, we conclude that the full subquiver $Q_0$ is the quiver of the $G^\prime$ theory coupled to a $R^\prime$ half--hyper. Repeating the procedure, we further reduce the gauge group. Eventually we end up with a theory with gauge group of the form $SU(2)^k$, for some $k$, coupled to half--hypers in suitable representations. This last theory is  \emph{complete} \cite{CV11}, and we know its quiver mutation--class from the classification of ref.\!\cite{CV11}. 

Therefore, if we are able to engineer a chain of Higgs decouplings, all associated to positive $\lambda_A(\cdot)$'s, which connects the theory of interest to an $SU(2)^k$ model, we produce a full subquiver $Q_0$ of the unknown quiver $Q$ of the our theory which is complete and hence explicitly known.

Our basic idea is to `induce' the unknown quiver $Q$ from its complete full subquivers $Q_0$. 
In facts, we concentrate on the \emph{maximal} complete subquivers, since knowing a larger subquiver reduces the guesswork required to reconstruct $Q$.

To induce $Q$ from its complete subquivers, we have to invert the Higgs decoupling map from $Q$ to $Q_0$. 
In the simplest case, eqn.\eqref{simplecase}, we give a large mass to just one simple--root $W$--boson, \textit{i.e.}\! the $a_0$--th one. The resulting gauge group $G^\prime$ will have (ignoring IR decoupled photons) rank $r(G^\prime)=r-1$, and hence the subquiver $Q_0$ must have \textit{at least} $2\, r(G^\prime)\equiv 2(r-1)$ nodes, corresponding to the linearly independent electric and magnetic weights of $G^\prime$. On the other hand, $Q_0$ has at least two less nodes\footnote{\ Indeed, since $Z(\delta_{a_0})\rightarrow\infty$, at least one node in the support of $\delta_{a_0}$ should become infinitely massive; also the nodes $e_i$ such that $m_{a_0}(e_i)\neq 0$ should become infinitely massive. Since $m_{a_0}(\delta_{a_0})=0$, if there is a node in $\mathrm{supp}(\delta_{a_0})$ with positive $m_{a_0}$--charge there must be also one with negative $m_{a_0}$--charge and both become infinitely massive; if all nodes in $\mathrm{supp}(\delta_{a_0})$ have zero $m_{a_0}$--charge,then there must be some other node with $m_a(e_i)\neq 0$ because the $m_{a_0}$--charge cannot be identically zero. In all cases, at least two nodes become infinitely massive and decouple.} than $Q$. 
Thus $Q_0$ has precisely two nodes less than $Q$. The $2(r-1)$ nodes of $Q_0$  must carry linearly independent electric/magnetic $G^\prime$ weights. Then, again, $\det B_0\neq 0$, and the theory which residues after decoupling has no conserved flavor charge, hence no `entire' hypermultiplets.

\begin{figure}
\begin{equation*}
\xymatrix{ \text{\fbox{\;$E_7\phantom{\Big|}$ $\tfrac{1}{2}\,\mathbf{56}$\;}} & \text{\begin{small}$SU(2)\times SO(12)$\:  $\tfrac{1}{2}(\mathbf{2},\mathbf{12})$\end{small}}\\
 \text{\fbox{\;$SO(12)\phantom{\Big|}$ $\tfrac{1}{2}\,\mathbf{32}$\;}}\ar@{=>}[u] & 
\text{\begin{small}$SU(2)\times SO(10)$\:  $\tfrac{1}{2}(\mathbf{2},\mathbf{10})$\end{small}}\ar@{..>}[ul]\ar@{..>}[u]\\
\text{\fbox{\;$SU(6)\phantom{\Big|}$ $\tfrac{1}{2}\,\mathbf{20}$\;}}\ar@{=>}[u] &
\text{\fbox{\;$SU(2)\times SO(8)\phantom{\Big|}$ $\tfrac{1}{2}\,(\mathbf{2},\mathbf{8})$\;}}\ar@{=>}[ul]\ar@{..>}[u]\\
\text{\fbox{\;$SU(2)\times SU(4)\phantom{\Big|}$ $\tfrac{1}{2}\,(\mathbf{2},\mathbf{6})$\;}}\ar@{=>}[u]\ar@{=>}[ur]\\
\text{\fbox{\;$SU(2)^3\phantom{\Big|}$ $\tfrac{1}{2}\,(\mathbf{2},\mathbf{2},\mathbf{2})$\;}}\ar@{=>}[u]
}
\end{equation*}
\caption{The chain of inverse Higgs decouplings we use to induce the quiver of the various theories starting from the known one for the $SU(2)^3$ model. The double arrow stands for an elementary step in the chain which increases the rank of the gauge group by $1$. Dashed lines correspond to purely formal constructions (the unboxed theories are not UV complete).}
\label{mapofinduction}
\end{figure}
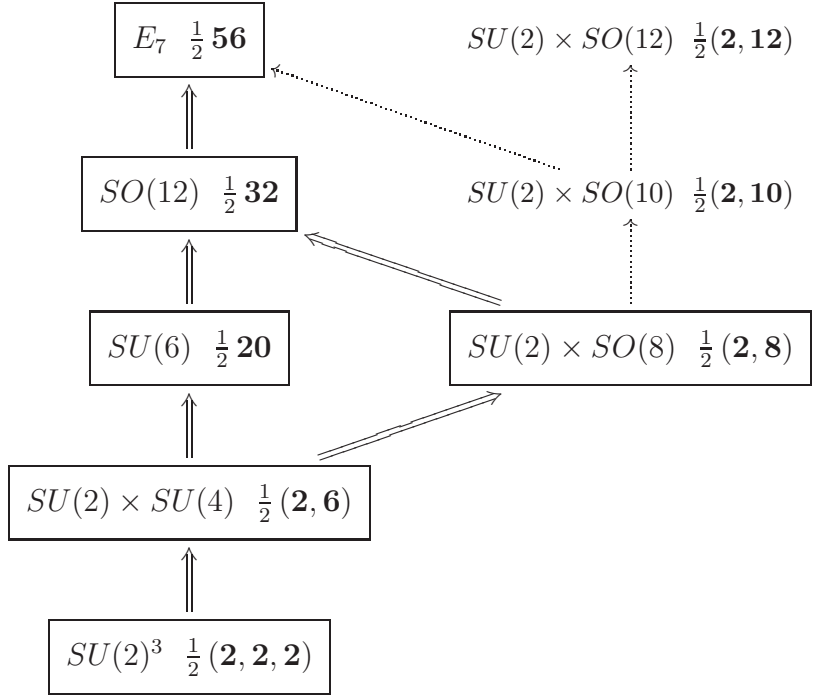

By considering the dual decoupling function 
\begin{equation}\widetilde{\lambda}(e_i)\equiv\begin{cases} 1 & \text{if }\lambda(e_i)=0\\
                                               0 & \text{otherwise},
                                              \end{cases}\end{equation}
which corresponds to the Higgs decoupling $G\rightarrow SU(2)\times U(1)^{r-1}$, we see that the complementary full subquiver $\widetilde{Q}_0$ on the two nodes with $\lambda(e_i)\neq 0$, must have a one--parameter family of representations corresponding to the light $SU(2)$ $W$--boson. This is possible if and only if these two massive nodes are connected precisely by two arrows, thus forming a Kronecker subquiver $\tau \rightrightarrows \omega$.

In conclusion, in the case of eqn.\eqref{simplecase}, to get the quiver $Q$ of the undecoupled theory one has only to specify the arrows connecting the two nodes $\tau$, $\omega$ of the massive Kronecker subquiver to the nodes $s\in Q_0$ of the known subquiver of the decoupled model. Indeed, except for $B_{\tau s}$, $B_{\omega s}$, all entries of the exchange matrix $B_{ij}$ of $Q$ are known. The Dirac integrality conditions state that the unknown entries $B_{\tau\, s}$, $B_{\omega\, s}$ should be such that the \textsc{rhs} of eqn.\eqref{magneticxxx} are \emph{integral} linear forms on $\Gamma$. In that equation the only unknowns are the integers $B_{\tau\, s}$, $B_{\omega\, s}$, since the $(\delta_b)_j$ are known being the charge vectors of the simple--roots of $G^\prime$, with support in $Q_0$, together with the minimal imaginary root $\delta_{a_0}$ of the massive Kronecker subquiver (seen as an affine $\widehat{A}_1$ Dynkin graph). In all examples below the Dirac integrality condition is strong enough to uniquely determine $B_{\tau\, s}$, $B_{\omega\, s}$, and hence the quiver $Q$.
\smallskip

Therefore, our strategy in this paper will be to start with a convenient complete theory, namely the $SU(2)^3$ gauge theory coupled to $\tfrac{1}{2}(\mathbf{2},\mathbf{2},\mathbf{2})$, and add just \emph{one} simple--root $W$--boson at each step (that is, we increase recursively the rank of $G$ by 1), by inverting the Higgs chain we described above. Eventually we shall arrive at the theories of interest, such as $E_7$ SYM coupled to $\tfrac{1}{2}\, \mathbf{56}$. In the process we construct several intermediate $\cn=2$ gauge theories coupled to half--hypers which are of independent interest. See figure \ref{mapofinduction}.

\subsubsection{Determing the superpotential $\cw$}

However, knowing $Q$ is not enough, one needs also the superpotential $\cw$. The terms of the superpotential involving only arrows of the decoupled full subquiver $Q_0$ are the same as in the superpotential $\cw_0$ of the $G^\prime$ theory and hence known. The other terms may be fixed (in principle) in two different ways:

The first technique \cite{cattoy} (which is the one we shall use) is to require that in all weak coupling chambers the light BPS states (that is, the BPS particles with zero magnetic charges) are vectors forming one copy of the adjoint of $G$ plus finitely many hypermultiplets in the correct representations of $G$ (this condition is called the `Ringel property' in ref.\!\cite{cattoy}). Note that a superpotential with this property must exist for the models of interests, since we know \emph{a priori} that they are quiver theories. The light BPS states at weak coupling may be analyzed using the techniques introduced in ref.\,\cite{cattoy}; the lowest order terms in $\cw-\cw_0$ are essentially universal, and may be read directly from \cite{cattoy}. The higher order ones may be fixed by educated trial and error. Again, they appear to be essentially unique (modulo isomorphism). \smallskip

The second technique to determine $\cw$ is the strong version of the $2d/4d$ correspondence. See refs.\,\cite{cattoy,arnold} for a discussion.

\subsection{The class of theories we focus on}\label{whichtheories?}

 We limit ourselves to sensible QFT's, namely the ones which are asymptotically free or UV conformal in \textit{all} gauge couplings and free of $\Z_2$--anomalies. Moreover, we restrict to gauge groups $G$ which are products of simply--laced simple Lie groups, just because these groups are best understood at the level of pure $\cn=2$ SYM. To avoid unnecessary complications, we also assume that there are no \emph{full}--hypermultiplets in the model; if necessary, they may be easily added by the `massive quark' procedure of \cite{CV11,ACCERV2}. The matter representation content is then of the form
\begin{equation}\label{const}
\begin{aligned}
&\frac{1}{2} \bigoplus_i\left(\bigotimes_\alpha R_{\alpha,i}\right)&&\text{where $R_{\alpha,i}$ is an irrepr.\! of the $\alpha$--th simple factor of $G$ }\\
& && \text{and } \bigotimes_\alpha R_{\alpha,i}\not\simeq
\bigotimes_\alpha R_{\alpha,j}\ \text{for }i\neq j.
\end{aligned}
\end{equation}
$\cn=2$ supersymmetry requires the irreducible $G$--representations\footnote{\ In particular, the irrepr.\! $\bigotimes_\alpha R_{\alpha,i}$ are non--trivial, since the trivial representation is real not quaternionic.} $\bigotimes_\alpha R_{\alpha,i}$ to be quaternionic. By Frobenius--Schur, this is equivalent to the condition that, for each $i$, an \emph{odd} number of the $R_{\alpha,i}$'s are irreducible \textit{quaternionic}, whereas all other $R_{\alpha,i}$'s are irreducible \textit{real}. The case of interest for us is the one in which there is  precisely one half--hyper in the spectrum in a representation \eqref{const}. 

\medskip

Under the above conditions, a gauge theory of a single half--hyper in a representation \eqref{const} is UV complete iff
\begin{equation}\label{UVcompl}
\forall\:\alpha\colon\ - 2 h_{\alpha} ^{\lor} +  \frac{1}{2} \sum_i \Delta(R_{\alpha,i})\,\prod_{\beta\neq \alpha}\dim R_{\beta,i}\leq 0,
\end{equation}
where $h_{\alpha} ^{\lor}$ is the dual Coxeter number of the $\alpha$-th simple factor of $G$, and $\Delta(R_{\alpha,i})$ is the Dynkin index of the irrep $R_{\alpha,i}$.

\section{The complete case: $G=SU(2)^k$}

Our strategy to construct the quivers for general (simply--laced) gauge groups $G$ is to induce them from their maximal complete subquivers (cfr.\! \S.\,\ref{strategy}). Consequently we begin by reviewing the half--hypers in the complete case following refs.\!\cite{CV11,cattoy}. The reader may prefer to jump ahead to \S.\,4.

\subsection{$SU(2)^k$ coupled to half--hypers as gentle $\cn=2$ models}

Asymptotically--free Lagrangian $\cn=2$ theories with gauge group $G=SU(2)^k$ are, in particular, complete quiver theories in 
the sense of \cite{CV11}.
The quiver mutation--classes of all such theories are described in \cite{CV11}. Except for $11$ exceptional cases, the mutation--finite classes arise \cite{triangulation1,felikson} from ideal triangulations of surfaces $\mathcal{C}_{g,n,b,\{c_i\}}$ with genus $g$, $n$ punctures, and $b$ boundary components, the $i$--th boundary having $c_i\geq 1$ marked points.

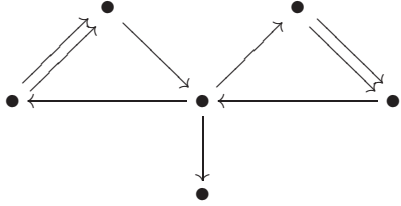
\begin{figure}
 \begin{equation*}
  \begin{gathered}
\xymatrix{& \bullet\ar[rd] && \bullet\ar@<0.4ex>[dr]\ar@<-0.4ex>[dr]  &\\
\bullet \ar@<0.4ex>[ur]\ar@<-0.4ex>[ur] && \bullet \ar[ll]\ar[ur]\ar[d] && \bullet\ar[ll]\\
&& \bullet &&}\end{gathered}
 \end{equation*}
\caption{The $X_6$ quiver corresponding to $SU(2)^3$ SYM coupled to $\tfrac{1}{2}(\mathbf{2},\mathbf{3},\mathbf{1})\oplus \tfrac{1}{2}(\mathbf{2},\mathbf{1},\mathbf{3})$.}
\label{xxxx6666}
\end{figure}

\medskip

The exceptional $\cn=2$ quiver theories were analized in \cite{CV11}. Only one of them belongs to the class of \S.\,\ref{whichtheories?}: the model associated to the Derksen--Owen $X_6$ quiver \cite{derksen}, see figure \ref{xxxx6666}. $X_6$ describes the $SU(2)^3$ gauge theory coupled to a $\tfrac{1}{2}(\mathbf{2},\mathbf{3},\mathbf{1})\oplus \tfrac{1}{2}(\mathbf{2},\mathbf{1},\mathbf{3})$.

\medskip

All other complete theories of the class in \S.\,\ref{whichtheories?} arise from triangulations of surfaces. The number of nodes of the quiver of a triangulation of $\cc_{g,n,b,\{c_i\}}$ is
\begin{equation}\label{rkQ}
m=6 g - 6 + 3 n + \sum_{i = 1} ^b (c_i + 3).
\end{equation}
We are interested in the complete theories with no flavor charges. The number of flavor charges is equal \cite{triangulation1} to $n$ plus the number of even $c_i$'s. However, \cite{CV11,cattoy} a boundary component with $c_i>1$ represents a $SU(2)$ SYM sector gauging the flavor symmetry of an Argyres--Douglas system of the $D_{c_i}$ type. Hence, to restrict ourselves to the complete theories of the class specified in section \ref{whichtheories?}, we have to set
\begin{equation}
 n=0,\quad c_i=1\ \ \text{for all }i.
\end{equation}
The number of $SU(2)$ gauge factors is
\begin{equation}\label{rcomp}
k \equiv m/2= 3g - 3 + 2 b.
\end{equation}
The surface $\mathcal{C}_{0,2}$ corresponds to pure $SU(2)$ SYM. All other $\mathcal{C}_{g,b}$ surfaces correspond to $SU(2)^k$ SYM coupled to half--hypermultiplets. Indeed, take a pant decomposition\footnote{\ Of course the pants decomposition is not unique. Different decompositions corresponds to different weakly coupled Lagrangian descriptions which are related by a web of Gaiotto $\cn=2$ dualities \cite{Gaiotto}.} of $\mathcal{C}_{g,b}$: in the corresponding degeneration limit 
each pair of pants produces a half--hyper in the $(\mathbf{2},\mathbf{2}, \mathbf{2})$ with respect to the three $SU(2)$ gauge factors associated to the long tubes attached to its three boundary components \cite{Gaiotto}. A special case happens when a long tube connects two boundary components of the same pair of pants, thus gauging the diagonal subgroup of the $SU(2)$'s associated to the two boundaries. In this case the pants contributes a hypermultiplet in the $\tfrac{1}{2}(\mathbf{1},\mathbf{2})\oplus \tfrac{1}{2}(\mathbf{3},\mathbf{2})$ representation of the $SU(2)_\mathrm{diag}\times SU(2)_\mathrm{3^\mathrm{rd}\;boundary}$ gauge subgroup.
\medskip

A convenient recipe to produce explicit quivers in each of the classes $\mathcal{C}_{g,b}$ can be found in  section 3 of \cite{ladkan1}. For quivers arising from triangulations of surfaces $\mathcal{C}_{g,b}$ the total number of arrows $\# Q_1 = 12(g-1) + 7 b$ is an invariant under mutations, a feature of all $SU(2)^k$ half--hypers models, but $X_6$ \cite{ladkan2}. Since $\mathcal{C}_{g,b}$ surfaces have no punctures, the associated potentials are just sum of oriented triangles, and the resulting Jacobian algebras $\C Q / (\partial \cw)$ are \emph{gentle} algebras \cite{assemgentle}\!\cite{cattoy}. In particular, this means that such algebras are \emph{tame}, so the BPS spectrum consists only of hypermultiplets and vector--multiplets, all higher spin states being forbidden \cite{cattoy}. The Representation Theory of the gentle algebras arising from triangulations of bordered surfaces is described in \cite{assemgentle}\!\!\cite{cattoy}:  It can be stated as the chain of correspondences
\begin{equation}\label{gentleasy}
\text{indecomposable modules } \longleftrightarrow \text{ strings/bands } \longleftrightarrow \text{ homotopy classes of curves}.
\end{equation}
The resulting description of the BPS states in terms of curves on $\cc_{g,b}$ coincides with the one obtained from geometric engineering \cite{Shapere:1999xr} and the WKB methods \cite{GMN09,ACCERV1}. 

\smallskip

For additional details about the $\cn=2$   $\cc_{g,b}$ models (called \emph{gentle}) see ref.\!\cite{cattoy}.

\subsection{$SU(2)^3$ coupled to $\tfrac{1}{2}(\mathbf{2},\mathbf{2},\mathbf{2})$}\label{222}

The pair of pants ($3$--holed sphere) $\cc_{0,3}$ is the basic building block of all gentle $\cn=2$ theories. It is also the starting point of our inverse Higgs chain, as discussed at the end of \S.\,\ref{strategy}.
Physically, $\cc_{0,3}$ corresponds to $SU(2)^3$ SYM coupled to $\tfrac{1}{2}(\mathbf{2},\mathbf{2},\mathbf{2})$.
\medskip

The canonical ideal triangulation of the $3$--holed sphere is invariant under cyclic permutations of the three holes. The corresponding $\Z_3$--symmetric quiver with superpotential $\cw$ is \cite{ACCERV2,cattoy,assemgentle} 
\begin{equation}\label{C03}
\begin{split}
Q_{0,3} \equiv\begin{gathered}
 \xymatrix{*++[o][F-]{1} \ar[rr]^{H_1} && *++[o][F-]{2}\ar[dl]_{H_2}\\
 & *++[o][F-]{3}\ar[lu]_{H_3}&\\
 &  &\\
 *++[o][F-]{5}\ar[uuu]_{V_1}\ar[dr]^{h_3} & & *++[o][F-]{6}\ar[ll]_{h_1\qquad}\ar[uuu]_{V_2}\\
& *++[o][F-]{4}\ar[ur]^{h_2}\ar[uuu]_{V_3} &\\
}
\end{gathered}&\qquad
\cw = H_1 H_3 H_2 + h_3 h_1 h_2
\end{split}
\end{equation}
Via the correspondence \eqref{gentleasy} we can recognize immediately the vectors belonging to the light category $\mathscr{L}(Q_{0,3},\cw)$, which correspond to the three $W$--bosons of the theory. The corresponding band representations are physically identified with the cycles having vanishing length in the shrinking limit of the holes of $\cc_{0,3}$ \cite{cattoy}. They are the representations of the three Euclidean $\widehat{A}(3,1)$ full subquivers with dimension vector equal to the respective primitive imaginary roots, $\dim W_i^\lambda=\fd_i$ ($i=1,2,3$)\footnote{\ $\lambda\in\mathbb{P}^1$. All solid lines without labels correspond to the map $1$. },
\begin{equation}\label{wbosons11}
W_1^\lambda =  \begin{gathered}\xymatrix@R=0.6pc@C=0.6pc{\C \ar@{..>}[rr] && 0\ar@{..>}[dl]\\
& \C\ar[lu] &\\
&  &\\
\C\ar[uuu]\ar[dr] & & 0\ar@{..>}[ll]\ar@{..>}[uuu]\\
& \C\ar@{..>}[ur]\ar[uuu]_\lambda &}\end{gathered}, \qquad W_2 =\begin{gathered}\xymatrix@R=0.6pc@C=0.6pc{\C \ar[rr] && \C \ar@{..>}[dl]\\
& 0\ar@{..>}[lu] &\\
&  &\\
\C\ar[uuu]^\lambda\ar@{..>}[dr] & & \C\ar[ll]\ar[uuu]\\
& 0\ar@{..>}[ur]\ar@{..>}[uuu] &}\end{gathered},\qquad W_3= \begin{gathered}\xymatrix@R=0.6pc@C=0.6pc{0 \ar@{..>}[rr] && \C\ar[dl]\\
& \C\ar@{..>}[lu] &\\
&  &\\
0\ar@{..>}[uuu]\ar@{..>}[dr] & & \C\ar@{..>}[ll]\ar[uuu]_\lambda\\
& \C\ar[ur]\ar[uuu] &}\end{gathered}.
\end{equation}

The three $W$--bosons are mutually local, $\langle \fd_i, \fd_j\rangle_\text{Dirac}=0$, as they should be.
The magnetic charges $m_i(X)$ of a representation $X \in \mathsf{rep}(Q,\cw)$ are given by eqn.\eqref{magneticxxx}. 
Setting $n_j = (\text{dim }X)_j$, we obtain
\begin{equation}\label{magn222}
\begin{aligned}
m_1=n_5-n_1,\qquad
m_2=n_6-n_2,\qquad
m_3=n_4-n_3,
\end{aligned}
\end{equation}
which are all integral as required by the Dirac integrality condition.
The weak-coupling regime associated to shrinking the holes of $\cc_{g,b}$ is then given by \cite{cattoy}
\begin{equation}\label{wc222}
\begin{cases}
\arg Z_i = O(g_j^2), & i=1,2,3\\
\arg Z_i = \pi - O(g_j ^2), & i=4,5,6,
\end{cases}
\end{equation}
where $g_j$ ($j=1,2,3$) are the three Yang--Mills couplings of the $SU(2)$ gauge factors  \cite{cattoy}.
  
In ref.\!\cite{cattoy} it is proven that in all chambers in the regime \eqref{wc222}, there is precisely one BPS hypermultiplet per each of the following charges
\begin{align}\label{chv1}
&\tfrac{1}{2}\text{ dim}(\fd_1+\fd_2-\fd_3)\hskip 3.2mm=(1,0,0,0,1,0)\\
&\tfrac{1}{2} \text{ dim}(\fd_1-\fd_2+\fd_3)\hskip 3.2mm=(0,0,1,1,0,0)\\
&\tfrac{1}{2}\text{ dim}(-\fd_1+\fd_2+\fd_3) = (0,1,0,0,0,1)\\
&\tfrac{1}{2}\text{ dim}(\fd_1+\fd_2+\fd_3)\hskip 3.2mm=(1,1,1,1,1,1),
\label{chv4}\end{align}
which, together with their PCT conjugates, give the weights of the representation $(\mathbf{2},\mathbf{2},\mathbf{2})$ of the $SU(2)^3$ symmetry associated to the $W$--bosons \eqref{wbosons11}. Hence the matter consists precisely of one $(\mathbf{2},\mathbf{2},\mathbf{2})$ `half'--quark which is local with respect to the three $W$--bosons.
The three $W$--bosons and the $\tfrac{1}{2}(\mathbf{2},\mathbf{2},\mathbf{2})$ half--hyper are the only BPS particles whose masses remains bounded as $g_j\rightarrow 0$. More details on the model may be found in ref.\!\cite{cattoy}.

\section{Running the inverse Higgs procedure}\label{first2}

As discussed in \S.\,\ref{strategy}, in our procedure we increase the rank of the gauge group by one at each step, starting from the maximal complete subsector, until we reach the theory of interest. In the first two steps we construct theories with gauge group
$$G_n\equiv SU(2)\times SO(2n),$$
coupled to a half--hyper in the quaternionic representation $\tfrac{1}{2}(\mathbf{2},\mathbf{2n})$ which are of independent interest. The condition of no Landau pole restricts $n$ to the values $2$, $3$, and $4$. Since $SO(4)\simeq SU(2)\times SU(2)$, the case $n=2$ coincides with the complete model discussed in \S.\,\ref{222}.
Since $r(G_n)=n+1$, making $n\rightarrow n+1$ amounts to one elementary step in our inverse Higgs procedure.

\subsection{$SU(2)\times SO(6)$ coupled to $\tfrac{1}{2}\,(\mathbf{2},\mathbf{6})$}

The simplest non--complete $\cn=2$ theory of the class described in \S.\,\ref{whichtheories?} is $SU(2)\times SO(6) \simeq SU(2) \times SU(4)$ SYM coupled to a half--hyper in the $(\mathbf{2},\mathbf{6})$.

\subsubsection{Inducing the quiver}

\begin{figure}
\begin{center}
 \begin{equation*}
  \xymatrix{
  *++[o][F-]{{\a}_{0}}&&*++[o][F-]{{\a}_1}\ar@{-}[r] & *++[o][F-]{{\a}_2} \ar@{-}[r] & *++[o][F-]{{\a}_3}}
 \end{equation*}
 \end{center}
 \caption{The Dynkin diagram of $SU(2)\times SU(4)$.}\label{26dyn}
 \end{figure}
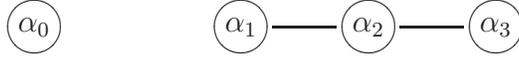

We write $\alpha_0$ and $\alpha_1,\alpha_2,\alpha_3$ for the simple roots of $\mathfrak{su}(2)$ and $\mathfrak{su}(4)$, respectively (see figure \ref{26dyn}). Following \S.\,\ref{strategy}, we consider a v.e.v.\! of the complex adjoint scalar $\langle\Phi\rangle\in\mathfrak{h}$ such that
\begin{equation}\label{rrrwww231}
 \alpha_a(\langle\Phi\rangle)= i\,t\,\delta_{a\,2}+O(1),\qquad t\rightarrow +\infty.
\end{equation}
The limit \eqref{rrrwww231} produces a Higgs decoupling corresponding to the breaking
\begin{equation}
 SU(2)\times SU(4)\to SU(2)^3\times U(1),
\end{equation}
where the $U(1)$ generator is $q =\text{diag}(1,1,-1,-1)\in\mathfrak{su}(4)$. The $(\mathbf{2},\mathbf{6})$ decomposes under the resulting gauge group
 as:
\begin{equation}
(\mathbf{2},\mathbf{6})\to (\mathbf{2},\mathbf{2},\mathbf{2})_0\oplus(\mathbf{2},\mathbf{1},\mathbf{1})_2\oplus(\mathbf{2},\mathbf{1},\mathbf{1})_{-2}.
\end{equation}
By the BPS bound, the states with finite mass as $t \rightarrow \infty$ must have $q=0$. Therefore, the decoupled theory is $SU(2)^3$ SYM coupled to a $\tfrac{1}{2}(\mathbf{2},\mathbf{2},\mathbf{2})$ hyper (plus an IR--free massless photon).

According to \S.\,\ref{strategy}, in the mutation--class of $SU(2) \times SU(4)$ SYM coupled to a $\tfrac{1}{2}(\mathbf{2},\mathbf{6})$ we may choose a quiver $Q$ which contains the quiver
\eqref{C03} as a \emph{full} subquiver $Q_0$. The complementary subquiver is then a Kronecker one, $\mathbf{Kr}$.
To get $Q$, it remains to fix the arrows connecting the two complementary full subquivers $Q_0$ and $\mathbf{Kr}$.
\smallskip

The $\mathbb{P}^1$--family of representations supported on the Kronecker subquiver,
$W_K^\lambda = \xymatrix{\C  \ar@<0.5ex>@{->}[r]^{\lambda}\ar@<-0.5ex>@{->}[r]_1 & \C}$,
 represents a $W$-boson supermultiplet which, according to the discussion in \S.\,\ref{strategy}, is associated to the simple root $\alpha_2$ (cfr.\! eqn.\eqref{rrrwww231}). 
The representations $W_i^\lambda\in\mathsf{rep}(Q_0,\cw_0)$, associated to the three $W$--bosons of the $SU(2)^3$ theory, are given in eqn.\eqref{wbosons11}. From figure \ref{26dyn} we see that they correspond to the $SU(2)\times SU(4)$ light $W$--bosons associated to the simple--roots $\alpha_0,\alpha_1$, and $\alpha_3$. We write $\fd_K\equiv\dim W^\lambda_K\in \Gamma$ and $\fd_i\equiv \dim W_i^\lambda$. 

The arrows connecting the two subquivers are determined by requiring that $i)$ the simple $W$--bosons are mutually local, and $ii)$ the corresponding magnetic charges \eqref{magneticxxx}
are integral forms. Choosing the connecting arrows in the following way
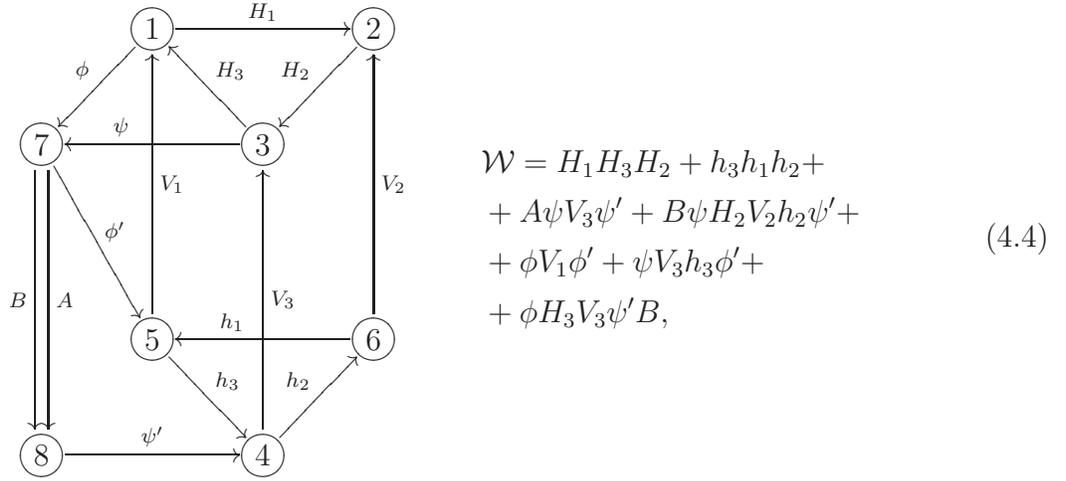
\begin{equation}\label{quiv26}
\begin{split}
\begin{gathered}
\xymatrix{&*++[o][F-]{1} \ar[ld]_{\phi} \ar[rr]^{H_1} && *++[o][F-]{2}\ar[dl]_{H_2}\\
*++[o][F-]{7} \ar@<0.5ex>@{->}[ddd]^{A}\ar@<-0.5ex>@{->}[ddd]_{B} \ar[ddr]^{\phi^{\prime}}& & *++[o][F-]{3}\ar[lu]_{H_3} \ar[ll]_{\psi \qquad} &\\
& &  &\\
 &*++[o][F-]{5}\ar[uuu]_{V_1}\ar[dr]^{h_3} & & *++[o][F-]{6}\ar[ll]_{h_1\qquad}\ar[uuu]_{V_2}\\
*++[o][F-]{8} \ar[rr]^{\psi^{\prime}}& & *++[o][F-]{4}\ar[ur]^{h_2}\ar[uuu]_{V_3} &\\
}
\end{gathered}&\qquad
\begin{aligned}
&\cw = H_1 H_3 H_2 + h_3 h_1 h_2 +\\
&+ A \psi V_3 \psi^{\prime} + B \psi H_2 V_2 h_2 \psi^{\prime} +\\
& + \phi V_1 \phi^{\prime} + \psi V_3 h_3 \phi^{\prime} +\\
& + \phi H_3 V_3 \psi^{\prime} B,
\end{aligned}
\end{split}
\end{equation}
we obtain the Dirac pairings (here $n_i = (\text{dim }X)_i$)
\begin{equation}\label{26dirac}
\begin{split}
&-\langle \fd_1, \text{dim }X \rangle_\text{Dirac} = 2 (n_5 -n_1) + (n_8 - n_7),\\
&-\langle \fd_2, \text{dim }X \rangle_\text{Dirac} = 2 (n_6 - n_2),\\
&-\langle \fd_3, \text{dim }X \rangle_\text{Dirac} = 2 (n_4 -n_3) + (n_8 - n_7),\\
&-\langle \fd_K, \text{dim }X \rangle_\text{Dirac} = 2 (n_7 -n_8) + (n_3 - n_4) + (n_1 - n_5).
\end{split}
\end{equation}
Requirement $ii)$ is then satisfied under our identification of the families of representations associated to each simple--root $W$--boson; in terms of dimension vectors, we have (cfr.\! fig.\,\ref{26dyn})
\begin{equation}\label{ttttw45}
 (\alpha_0,\alpha_1,\alpha_2,\alpha_3)\longrightarrow (\delta_0,\delta_1,\delta_2,\delta_3)\equiv (\fd_2,\fd_1,\fd_K,\fd_3).
\end{equation}
which give
the integral magnetic charges
\begin{equation}\label{magn26}
\quad m_0 =n_6 - n_2, \quad m_1 = n_5 - n_1, \quad m_2 = n_7 - n_8,\quad m_3 =  n_4 - n_3.
\end{equation}

This completes the induction of the quiver $Q$ from its maximal complete subquiver $Q_0\equiv Q_{0,3}$.
In eqn.\eqref{quiv26} we wrote also the corresponding superpotential $\cw$ as determined from the considerations in the next subsection. With this superpotential, the Jacobian algebra $\C Q/(\partial\cw)$ is finite dimensional (see appendix \ref{app:finite}).

\subsubsection{Light BPS particles at weak coupling}\label{light26}

Physically, the most convincing proof that the pair $(Q,\cw)$ in eqn.\eqref{quiv26} do correspond to $SU(2)\times SU(4)$ SYM coupled to $\tfrac{1}{2}(\mathbf{2},\mathbf{6})$ is to extract from $\mathsf{rep}(Q,\cw)$ the BPS spectrum in the limit
$g_2,g_4\rightarrow 0$, where $g_2,g_4$ are the two Yang--Mills couplings. A part for dyons with masses $O(1/g^2)$, in this limit  the spectrum must coincide with the perturbative one: vector--multiplets making one copy of the adjoint of $SU(2)\times SU(4)$ plus hypermultiplets in the $\tfrac{1}{2}(\mathbf{2},\mathbf{6})$ representation.
The light perturbative states are the stable objects of the light  subcategory $\mathscr{L}(Q,\cw)$ 
defined in section
\ref{magneticweights} (see \cite{cattoy} for details).
An object $X \in \mathsf{rep}(Q,\cw)$ lies in $\mathscr{L}(Q,\cw)$ iff 
\begin{equation}
\begin{split}
&1)\, \, m_i(X) = 0, \, \, \forall \, \,i=1,2,3,K;\\
&2)\, \, m_i(Y) \leq 0,\hskip1.8mm \forall\ \text{subobject } Y\ \text{of }X.
\end{split}
\end{equation}
In particular, we may consider the light categories for the Kronecker quiver, $\mathscr{L}(\mathbf{Kr})$, which corresponds to the perturbative sector of pure $SU(2)$ SYM, and for the $\widehat{A}(3,1)$ affine quiver, $\mathscr{L}(\widehat{A}(3,1))$, corresponding   to the `perturbative' (in $g_\mathrm{YM}$) sector of $SU(2)$ SYM coupled to the $A_3$ Argyres--Douglas theory \cite{CV11,cattoy}.
\medskip

The Higgs mechanism may be consistently implemented in perturbation theory (assuming the gauge couplings to be small enough). This means that no perturbative state may become non--perturbative in the breaking process $SU(2)\times SU(4)\to SU(2)^3\times U(1)$ (or in the complementary one $SU(2)\times SU(4)\rightarrow SU(2)\times U(1)^3$). In terms of representations, this requires that if $X$ belongs to the perturbative category $\mathscr{L}(Q,\cw)$, so do its restrictions to the subquivers resulting from the Higgs decoupling. Then we must have
\begin{equation}\label{llllemmma}
X\in \mathscr{L}(Q,\cw)\quad\Longrightarrow\quad\left\{\begin{aligned}
&\;X \big|_{\mathbf{Kr}} \hskip5.5mm\in \mathscr{L}(\mathbf{Kr})\\
&\;X \big|_{\hat{A}(3,1)_i} \in \mathscr{L}({\widehat{A}(3,1)}), \quad \forall \, \, i=1,2,3.
\end{aligned}\right.
\end{equation}

The first check on the correctness of the pair $(Q,\cw)$ is that \eqref{llllemmma} holds true for the category $\mathsf{rep}(Q,\cw)$. 
The mathematical proof of eqn.\eqref{llllemmma} is presented in appendix \ref{lemma1}.
Eqn.\eqref{llllemmma} implies \cite{cattoy} that, for a light
representation $X$,  the arrows $V_i$ ($i=1,2,3$) and $B$ of the quiver \eqref{quiv26} are isomorphisms\footnote{\ \label{foott}Strictly speaking this is true over $N$ minus the two points at infinity in the two $\mathbb{P}^1$ irreducible components. Here $N$ is the index variety for $\mathscr{L}(Q,\cw)$ (see \cite{cattoy}). This subtlety is totally harmless, since there is no matter over the two omitted points. Its only effect is that one has to take the projective closure of the final answer.}. These isomorphisms allow to identificate the upper and lower nodes of the quiver \eqref{quiv26} in pairs. Replacing $V_1\rightarrow -1$ in $\cw$, the fields $\phi$, $\phi^\prime$ become massive and may be integrated away, producing a new term $\psi h_3H_3\psi^\prime$ in the effective superpotential. 
 We conclude that (with the qualification in footnote \ref{foott}) the `perturbative' category $\mathscr{L}(Q,\cw)$ for the pair in eqn.\eqref{quiv26} is equivalent to
 the category of the representations of the quiver
\begin{equation}\label{geom26}
Q^{\,\prime}\equiv
\begin{gathered}
\xymatrix@R=2.0pc@C=3.0pc{
&&&&1 \ar@/_0.5pc/[dll]_{h_3} \ar@/_0.5pc/[dd]_{H_1}\\
&\ar@(ul,dl)[]_{A} 0 \ar@/_0.7pc/[r]_{\psi^{\prime}} & 3 \ar@/_0.7pc/[l]_{\psi} \ar@/_0.5pc/[urr]_{H_3} \ar@/_0.5pc/[drr]_{h_2}&&\\
&&&&2 \ar@/_0.5pc/[ull]_{H_2} \ar@/_0.5pc/[uu]_{h_1}
}
\end{gathered}
\end{equation}
subjected to the relations $\partial\cw^\prime=0$ 
which follow from the effective superpotential
\begin{equation}\label{redSQM26}
\cw^{\,\,\prime} = H_1 H_3 H_2 + h_3 h_1 h_2 + A \psi \psi^{\prime} + \psi H_2 h_2 \psi^{\prime} + \psi h_3 H_3 \psi^{\prime}.
\end{equation}

In a weakly coupled regime,
\begin{equation}\label{eeeexxx}
\begin{cases}
 \arg Z_i= O(g_2^2,g_4^2) & i=1,2,3,8\\
\arg Z_i= \pi -O(g_2^2, g_4^2) & i=4,5,6,7, 
\end{cases}
\end{equation}
the
stable representations of $\mathscr{L}(Q,\cw)$ are in one--to--one correspondence with the stable objects of $\mathsf{rep}(Q^{\,\prime},\cw^{\,\prime})$ with respect to the induced central charge $Z$.
This description of $\mathscr{L}(Q,\cw)$ is parallel to the one given in \cite{cattoy} for $\cn=2$ SYM coupled to massive hypermultiplets.

\medskip

In \cite{cattoy} it is shown that if
\begin{equation}
 \langle\cdot,\cdot\rangle_\text{Dirac}\Big|_{\mathscr{L}(Q,\cw)}\equiv 0,
\end{equation}
  the set $\{\gamma\}$ of the charge vectors of the light BPS states at weak coupling (that is, the set of the dimension vectors of the $Z$--stable objects of $\mathscr{L}(Q,\cw)$) is independent of the particular chamber, as long as it is weakly coupled, (that is, it satisfies eqn.\eqref{eeeexxx}). The actual stable representations, however, \textit{do} depend on the chamber \cite{cattoy}. This means that the set of dimensions of the stable objects of  $\mathsf{rep}(Q^{\,\prime},\cw^{\,\prime})$
is independent of the induced central charge $Z$. Therefore, to determine these dimensions, which give the quantum numbers of the light BPS particles at weak coupling, we may choose any particular chamber we like. 
We find convenient to work in the chamber specified by the induced central charge
\begin{equation}\label{chmb26}
\begin{aligned}
&Z_0 \text{ real positive and } |Z_0| \ggg |Z_1|,|Z_2|,|Z_3|\\
&0 < \text{arg }Z_1, \text{arg }Z_2, \text{arg }Z_3.
\end{aligned}
\end{equation}

We focus on representations\footnote{\ Given a representation $X$, we write $X_i$ for the vector space associated to the $i$--th node of $Q^\prime$.}  $X$ of the pair $(Q^\prime,\cw^\prime)$
 having $X_0\neq 0$, since the ones with $X_0=0$ come from representations of the quiver $Q_{0,3}$ (eqn.\eqref{C03}) already studied in \S.\,\ref{222}.
A stable representation with $X_0\neq0$ is either the image of an element of the
 $\mathbb{P}^1$--family of the $W_K$-- boson, or has $X_3\neq 0$. Hence we may assume both spaces $X_0,X_3\neq0$ to be not zero.
\smallskip

The subrepresentation\footnote{\ For a check that \eqref{subspacesY} is indeed a subrepresentation, see appendix \ref{pppqqw}.}
\begin{equation}\label{subspacesY}0\rightarrow \big(0,\text{Im}(H_3\psi^\prime\psi),
\text{Im}(h_2\psi^\prime\psi),\text{Im}(\psi^\prime\psi)\big)\rightarrow (X_0,X_1,X_2,X_3)\end{equation}
is destablizing in the chamber \eqref{chmb26} unless it vanishes. Thus, for a stable representation $X$ with $X_0,X_3\neq 0$
\begin{equation}
 \psi^\prime\psi=0.\label{rrreeep}
\end{equation}
In facts, since (for stability) $\psi$ must be surjective, \eqref{rrreeep} implies the stronger condition $\psi^\prime=0$. 
It is easy to check that for $X\in\mathsf{rep}(Q^\prime,\cw^\prime)$ the following linear map
\begin{equation}
\ell\colon (X_0,X_1,X_2,X_3) \longmapsto \big(A X_0, - H_3 h_3 X_1, -h_2 H_2 X_2, -(H_2 h_2 + h_3 H_3)X_3 \big)
\end{equation}
is an element of End $X$. Since a stable representation is, in particular, a brick \cite{king,keller}, for $X$ stable $\ell$ should be a number $-\lambda\in\mathbb{C}$. Then the arrows of $X$ satisfy the relations
\begin{equation}
H_3h_3=h_2H_2=H_2h_2+h_3H_3=\lambda\quad \big(=-A\big).\label{rere2}
\end{equation}

$X$ is, in particular, a representation of the (non--full) subquiver $Q_\text{forget}$ obtained by forgetting the arrows $h_1,H_1$ and $A$ in the quiver $Q^\prime$ of eqn.\eqref{geom26} 
\begin{equation}\label{hhhhh1}
Q_\text{forget}\:\equiv\hskip-6mm\begin{gathered}
\xymatrix@R=2.0pc@C=2.0pc{
&&&&&1 \ar@/_0.5pc/[dll]_{h_3}\\
& 0 \ar@/_0.5pc/[rr]_{\psi^{\prime}} && 3 \ar@/_0.5pc/[ll]_{\psi} \ar@/_0.5pc/[urr]_{H_3} \ar@/_0.5pc/[drr]_{h_2}&&\\
&&&&&2 \ar@/_0.5pc/[ull]_{H_2}
}
\end{gathered}
\end{equation}
(One can show \emph{a priori}, see appendix \ref{pppqqw}, that $h_1=H_1=0$ for all stable representations with $X_0,X_3\neq0$. So,  actually, we are not forgetting anything in passing from $Q^\prime$ to $Q_\text{forget}$).
\smallskip

It is convenient to relabel the arrows of $Q_\text{forget}$. The three `directed' arrows $\psi,H_3,h_2$ will be denoted as $\alpha_k$, $k=0,1,2$ respectively (each arrow having the same number as its target), while the corresponding `reversed' arrows $\psi^\prime, h_3, H_2$ will be denoted as $\alpha^*_k$ (same numbering as their sources). Then $\partial_A\cw^\prime=0$ together with the equations \eqref{rrreeep}\eqref{rere2} imply, in particular, the relations
\begin{equation}\begin{gathered} \label{ptttgX}
\sum\nolimits_k \,\alpha^*_k\,\alpha_k= \lambda\\
\alpha_0\,\alpha_0^* =0\\
\alpha_1\,\alpha_1^*=\alpha_2\,\alpha_2^*=\lambda.\end{gathered}\end{equation}
There are two cases: for $\lambda=0$, $Q_\text{forget}$ with the relations \eqref{ptttgX} give the preprojective algebra $\mathcal{P}\!(D_4)$ \cite{pre1,pre2,CBlemma,GLS} (see \cite{cattoy} for a discussion from the physical side); for $\lambda\neq 0$ we get a \emph{deformed preprojective algebra} $\Pi^\lambda(D_4)$ in the sense of Crawley--Bovey and Holland \cite{defr1,defr2,defr3,rump}. $\Pi^\lambda(D_4)$ is finite--dimensional and independent of $\lambda$ up to isomorphism.
\smallskip

$\lambda\in \mathbb{P}^1$ should be identified with the parameter\footnote{\ Keeping into account the footnote \ref{foott}.} of the families of stable representations associated to the BPS vector--multiplet, so the representations of $\Pi^\lambda(D_4)$ correspond to vector--multiplets, while the bricks of the preprojective algebra $\mathcal{P}(D_4)$ which are not in the projective closure of the $\Pi^\lambda(D_4)$ ones, correspond to the matter category (that is, to hypermultiplets). An obvious necessary condition for the existence of a representation of $\Pi^\lambda(D_4)$ is
\begin{equation}\label{ddim}
\dim X_3=\dim X_1+\dim X_2.
\end{equation}
Now, we may further specialize the choice \eqref{chmb26} such that 
\begin{equation}
 \big(X_0/\psi(\text{Im }h_3+\text{Im }H_2),0,0, X_3/(\text{Im }h_3+\text{Im }H_2)\big)
\end{equation}
is a destabilizing quotient, so that stability requires $X_3=\text{Im }h_3+\text{Im }H_2$, which, in view of eqn.\eqref{ddim}, implies
\begin{equation}\label{kkkkk223451}
 X_3\simeq X_1\oplus X_2.
\end{equation}
 Together with $\psi^\prime=0$, this last condition guarantees that all relations $\partial\cw^\prime=0$ are satisfied. So all representations of the deformed preprojective algebra which are stable for the induced central charge \emph{do} correspond to BPS vector--multiplets. 
Moreover, eqn.\eqref{kkkkk223451} implies that all indecomposable representations of $\Pi^\lambda(D_4)$ with $\psi^\prime=0$ have dimension vectors which are positive roots of $D_4$, orthogonal to the deformation, that is, satisfying eqn.\eqref{ddim}. These indecomposable representations are simple, \cite{defr1,defr2,defr3,rump}, hence stable.  Rewriting the dimensions vectors in terms of the physical charge vectors, eqn.\eqref{ttttw45}, we conclude that, in any weakly coupled chamber, the BPS vector--multiplets which have a bounded mass in the $g_2,g_4\rightarrow 0$ limit are precisely one per each of the following  charge vectors
\begin{equation}
 \label{tttyyy2}
\delta_1,\ \delta_2,\ \delta_3,\ \delta_1+\delta_2,\ \delta_2+\delta_3, \delta_1+\delta_2+\delta_3,\quad \underline{\delta_0},
\end{equation}
where the underlined one is the dimension vector of the stable representations which, having $X_0=X_3=0$, are not covered by the above analysis. They are the representations $W_2^\lambda$ of the complete subquiver \eqref{C03}.
This last dimension vector corresponds to the $SU(2)$ $W$--boson (cfr.\! eqn.\eqref{ttttw45}):  Thus the seven vector multiplets have the correct weights (quantum numbers) to be the positive roots of $\mathfrak{su}(4)\oplus\mathfrak{su}(2)$.
The light vector--multiplets form one copy of the adjoint of $SU(2)\times SU(4)$, as physically required.

At $\lambda=0$, the stable representations are in one--to--one correspondence with the positive roots of $D_4$. Omitting those in the projective closure of the $\mathfrak{su}(4)$ (see footnote \ref{foott}), and focusing on the rigid ones, we get one BPS hypermultiplet per each of the following  physical charge vectors,
\begin{equation}\label{yyyyxxxbbp}
 \begin{aligned}
  &\tfrac{1}{2}(\delta_1-\delta_3+\delta_0) && \tfrac{1}{2}(-\delta_1+\delta_3+\delta_0) && \tfrac{1}{2}(\delta_1+\delta_3-\delta_0)\\
&\delta_2+\tfrac{1}{2}(\delta_1+\delta_3-\delta_0) && \tfrac{1}{2}(\delta_1+\delta_3+\delta_0) && \delta_2+\tfrac{1}{2}(\delta_1+\delta_3+\delta_0)
 \end{aligned}
\end{equation}
which, together with their PCT conjugates, give precisely the $SU(2)\times SU(4)$ weights\footnote{\ In our basis. To get the expression in the Dynkin basis one has to multiply by the Cartan matrix.} of a $\tfrac{1}{2}(\mathbf{2},\mathbf{6})$. This is hardly a surprise, since the quiver $Q_\text{forget}$ produces all the roots of $SO(8)$, and the adjoint of $SO(8)$ decomposes under the $SU(4)$ subgroup in the adjoint plus two $\mathbf{6}$.

This concludes the check that the pair $(Q,\cw)$ in eqn.\eqref{quiv26} produces the right perturbative spectrum as $g_2,g_4\rightarrow 0$. Our guessed $(Q,\cw)$ have been proven to be correct, and we may safely use the pair $(Q,\cw)$ for  non--perturbative computations and, in particular, to determine the BPS spectrum in the extreme strong coupling regime.

\subsection{$SU(2)\times SO(8)$ coupled to $\tfrac{1}{2}(\mathbf{2},\mathbf{8})$}

\begin{figure}
\begin{center}
 \begin{equation*}
  \xymatrix{
  && & &  *++[o][F-]{{\a}_3}\\
  *++[o][F-]{{\a}_{0}}&&*++[o][F-]{{\a}_1}\ar@{-}[r] & *++[o][F-]{{\a}_2} \ar@{-}[ur]\ar@{-}[dr] &\\
  &&&&  *++[o][F-]{{\a}_4}}
 \end{equation*}
 \end{center}\vglue-9pt
 \caption{The Dynkin diagram of $SU(2)\times SO(8)$.}\label{28dyn}
 \end{figure}
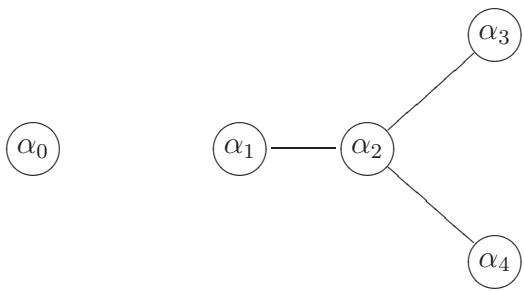

\subsubsection{Inducing the quiver}

The Dynkin graph of $SU(2)\times SO(8)$ is represented in figure \ref{28dyn}. The Higgs decoupling which gives back the previous $SU(2)\times SO(6)$ model corresponds to a background $\langle \Phi\rangle$ with
\begin{equation}
 \alpha_a(\langle\Phi\rangle)= i\,t\, \delta_{a\,1}+O(1),\quad t\rightarrow +\infty.
\end{equation}
The Higgs procedure of \S.\ref{strategy} then leads to the following pair $(Q,\cw)$
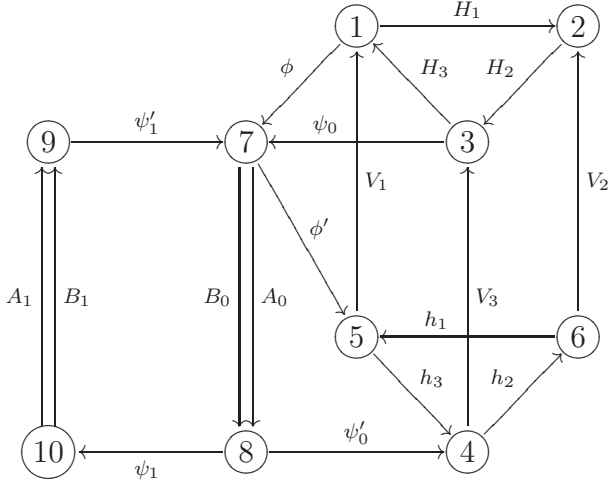
\begin{equation}\label{quiv28}
\begin{split}
\begin{gathered}
\xymatrix{&&&*++[o][F-]{1} \ar[ld]_{\phi} \ar[rr]^{H_1} && *++[o][F-]{2}\ar[dl]_{H_2}\\
*++[o][F-]{9}\ar[rr]^{\psi_1^{\prime}}&&*++[o][F-]{7} \ar@<0.5ex>@{->}[ddd]^{A_0}\ar@<-0.5ex>@{->}[ddd]_{B_0} \ar[ddr]^{\phi^{\prime}}& & *++[o][F-]{3}\ar[lu]_{H_3} \ar[ll]_{\psi_0 \qquad} &\\
&&& &  &\\
 &&&*++[o][F-]{5}\ar[uuu]_{V_1}\ar[dr]^{h_3} & & *++[o][F-]{6}\ar[ll]_{h_1\qquad}\ar[uuu]_{V_2}\\
*++[o][F-]{10} \ar@<0.5ex>@{->}[uuu]^{A_1}\ar@<-0.5ex>@{->}[uuu]_{B_1}&&*++[o][F-]{8}\ar[ll]^{\psi_1}  \ar[rr]^{\psi_0^{\prime}}& & *++[o][F-]{4}\ar[ur]^{h_2}\ar[uuu]_{V_3} &\\
}
\end{gathered}&\qquad
\begin{aligned}
&\cw = H_1 H_3 H_2 + h_3 h_1 h_2 +\\
&+ A_0 \psi_0 V_3 \psi_0^{\prime} + B_0 \psi_0 H_2 V_2 h_2 \psi_0^{\prime} +\\
& + \phi V_1 \phi^{\prime} + \psi_0 V_3 h_3 \phi^{\prime} +\\
& + \phi H_3 V_3 \psi_0^{\prime} B_0 + \\
&+ A_1 \psi_1 A_0 \psi_1^{\prime}  +  B_1 \psi_1 B_0 \psi_1^{\prime}\\
& = \cw_0 + A_1 \psi_1 A_0 \psi_1^{\prime}  +  B_1 \psi_1 B_0 \psi_1^{\prime},
\end{aligned}
\end{split}
\end{equation}
where $\cw_0$ is the superpotential of the full subquiver \eqref{quiv26} and all other terms are uniquely fixed by \cite{cattoy}.
This quiver produces the  \textit{integral} magnetic charges
\begin{equation}\label{magn28}
 m_0 =n_6 - n_2, \quad m_1 = n_4 - n_3,\quad m_{2} = n_7 - n_8, \quad m_3 =  n_5 - n_1,  \quad m_{4} = n_{10}-n_9.
\end{equation}
under the identifications of the simple--root $W$--boson representations implied by figure \ref{28dyn} (unique up to $SO(8)$ triality)
\begin{equation}\label{D5miracle}
(\alpha_0,\alpha_1,\alpha_2,\alpha_3,\alpha_4)\longrightarrow(\d_0,\d_1,\d_2,\d_3,\d_4)\equiv (\fd_2,\fd_1,\fd_{K_0},\fd_3,\fd_{K_1}). 
\end{equation}

As a  check on the correctness of the pair $(Q,\cw)$ we compute the light BPS states as $g_2,g_8\rightarrow 0$ and match it with the expected perturbative
spectrum.

\subsubsection{Light BPS states at weak coupling}\label{light28}
Again our quiver is consistent with perturbative Higgs decoupling (see appendix \ref{lemma1}),
so if $X$ belongs to the perturbative category $\mathscr{L}(Q,\cw)$, then
\begin{equation}\label{secondfl}
X\in \mathscr{L}(Q,\cw)\quad\Longrightarrow\quad\left\{\begin{aligned}
&\;X \big|_{\mathbf{Kr}_s} \hskip4.2mm\in \mathscr{L}(\mathbf{Kr})\hskip1.2cm \forall\,\, s=0,1\\
&\;X \big|_{\hat{A}(3,1)_i} \in \mathscr{L}({\widehat{A}(3,1)}), \quad \forall \, \, i=1,2,3.
\end{aligned}\right.
\end{equation}
Consequently, as in \S.\,\ref{light26}, we identify the nodes on the top part of the quiver \eqref{quiv28} with those on the bottom, and integrate out the massive fields. We conclude that the perturbative category $\mathscr{L}(Q,\cw)$ is equivalent to $\mathsf{rep}(Q^\prime,\cw^\prime)$ 
where
\begin{equation}\label{geom28}
\begin{split}
Q^{\,\prime}\equiv
\begin{gathered}
\xymatrix@R=2.0pc@C=3.0pc{
&&&&&1 \ar@/_0.5pc/[dll]_{h_3} \ar@/_0.5pc/[dd]_{H_1}\\
&\ar@(ul,dl)[]_{A_{-1}} -1 \ar@/_0.7pc/[r]_{\psi_{-1}^{\prime}} &\ar@/_0.7pc/[l]_{\psi_{-1}} \ar@(dr,dl)[]^{A_0} 0 \ar@/_0.7pc/[r]_{\psi_0^{\prime}} & 3 \ar@/_0.7pc/[l]_{\psi_0} \ar@/_0.5pc/[urr]_{H_3} \ar@/_0.5pc/[drr]_{h_2}&&\\
&&&&&2 \ar@/_0.5pc/[ull]_{H_2} \ar@/_0.5pc/[uu]_{h_1}
}
\end{gathered}\qquad
&\begin{aligned}
\cw^{\,\prime} &= \cw^{\,\prime} _0+\\
& + \psi_{-1} A_0 \psi_{-1}^{\prime}+\\
& + A_{-1} \psi_{-1} \psi^{\prime}_{-1},
\end{aligned}\\
\end{split}
\end{equation}
and $\cw^{\, \prime} _0$ is the potential \eqref{redSQM26}.
We choose the convenient chamber
\begin{align}\label{chmb28}
&\arg Z_{-1}=0,\quad \quad 0<\arg Z_0, \arg Z_1, \arg Z_2, \arg Z_3,\quad |Z_{-1}|\ \text{large}.
\end{align}
We focus on stable representations $X\in\mathsf{rep}(Q^\prime,\cw^\prime)$ with $X_{-1}, X_{0}, X_3\neq 0$, since otherwise we get either a representation already studied in \S.\,\ref{light26} or a known pure $SU(3)$ SYM representation \cite{ACCERV2,cattoy}.
The subrepresentation
\begin{equation}
 0\rightarrow \big(0,\mathrm{Im}(\psi_0\psi^\prime_0),\mathrm{Im}(H_3\psi^\prime_0\psi_0),\mathrm{Im}(h_2\psi^\prime_0\psi_0), \mathrm{Im}(\psi^\prime_0\psi_0)
\big)\rightarrow (X_{-1},X_{0}, X_1, X_2, X_3)\end{equation}
is destabilizing unless $\psi_0\psi_0^\prime=\psi^\prime_0\psi_0=0$, which imply $\psi^\prime_{-1}\psi_{-1}=0$. Then
the quotient $Y$ with $Y_{-1}= X_{-1}/\ker \psi^\prime_{-1}$ and $Y_{i\neq -1}=0$ is destabilizing unless $\ker \psi^\prime=X_{-1}$, that is,
$\psi^\prime_{-1}=0$.\smallskip

The following map is an element of End $X$
\begin{equation*}
(X_{-1},X_{0},X_1,X_2,X_3) \longmapsto (-A_{-1} X_{-1}, A_0 X_{0}, - H_3 h_3 X_1, -h_2 H_2 X_2, -(H_2 h_2 + h_3 H_3)X_3 ).
\end{equation*}
Since $X$ must be a brick, there is $\lambda\in \C$ such that
\begin{equation}\label{prep28}
A_{-1} = -A_0 = H_3 h_3 = h_2 H_2 =   H_2 h_2 + h_3 H_3=\lambda.
\end{equation}
Again, we are reduced to representation of the double $\overline{D_5}$ of the $D_5$ quiver
\begin{equation}\label{geom28for}
 \overline{D_5}\colon
\hskip-1cm\begin{gathered}
\xymatrix@R=2.0pc@C=3.0pc{
&&&&&1 \ar@/_0.5pc/[dll]_{h_3} \\
& -1 \ar@/_0.7pc/[r]_{\psi_{-1}^{\prime}} &\ar@/_0.7pc/[l]_{\psi_{-1}}  0 \ar@/_0.7pc/[r]_{\psi_0^{\prime}} & 3 \ar@/_0.7pc/[l]_{\psi_0} \ar@/_0.5pc/[urr]_{H_3} \ar@/_0.5pc/[drr]_{h_2}&&\\
&&&&&2 \ar@/_0.5pc/[ull]_{H_2}
}
\end{gathered}
\end{equation}
(one may simply `forget' the arrows $h_1, H_1$ as we did in \S.\ref{light26}; it is easy to show that they are actually zero for a stable representation with $X_{-1}, X_{0}, X_3\neq 0$). 

For $\lambda\neq 0$, the representation $X$ satisfies, in particular, the relations of the deformed preprojective algebra  $\Pi^\lambda(D_5)$ in the form\footnote{\ The $\ast$ denotes the map $\boldsymbol{\alpha}\rightarrow \boldsymbol{\alpha}^*$ which to any arrow of the double cover associates the \textit{inverse} arrow.}
\begin{equation}
 \sum_{t(\boldsymbol{\alpha})=i}\boldsymbol{\alpha}^*\boldsymbol{\alpha}=\begin{cases}
\lambda & i=1,2,3\\
0 & i=-1,0,
       \end{cases}
\end{equation}
while at $\lambda=0$ we get the undeformed preprojective algebra\footnote{\ \label{subtleeety} In the literature there are several distinct definitions of \emph{the} preprojective algebra of a quiver $Q$. The different definitions correspond to different choice of a sign function from the set of all arrows of $Q$ to $\{\pm1\}$, see \cite{pre2}. All these definitions lead to isomorphic algebra if $Q$ is a tree, but not in general. The definition of preprojective algebra used in \cite{defr1} (before deformation) makes the sign choice which allows to write all generating relations as commutators. In our framework, we get naturally the sign assignements which makes them \emph{anti}commutators. This is unmaterial here, since we have $h_1=H_1=0$ which allows to reduce $Q$ to a tree.} $\cp(D_5)$.

As in \S.\ref{light26}, the BPS vector--multiplets which remain light as $g_2,g_8\rightarrow 0$ have quantum numbers corresponding to the positive roots of $D_5$ orthogonal to the deformation, that is, such that
\begin{equation}
 \dim X_3=\dim X_1+\dim X_2,
\end{equation}
while all other roots of $D_5$ correspond to light BPS hypermultiplets. In addition there is the $SU(2)$ $W$--boson whose representations have $X_{-1},X_{0},X_3=0$, and (again) are not covered by the previous argument.
Explicitly, the vector--multiplets charge vectors are 
\begin{equation}\begin{aligned}
 &\delta_1, \quad\delta_2, \quad\delta_{3}, \quad\delta_{4}, \quad\delta_1+\delta_{2}, \quad\delta_3+\delta_{2}, \quad\delta_{4}+\delta_{2},\\
&\delta_1+\delta_2+\delta_{3}, \quad\delta_1+\delta_{2}+\delta_{4}, \quad \delta_3+\delta_{2}+\delta_{4},\quad \delta_1+\delta_2+\delta_{3}+\delta_{4},\\ & \delta_1+\delta_3+\delta_{4}+2\,\delta_{2}, \qquad\underline{\delta_0},
\end{aligned}\end{equation}
which, in view of eqn.\eqref{D5miracle}, are precisely the weights of the adjoint of $\mathfrak{so}(8)\oplus \mathfrak{su}(2)$.
The other 8 roots correspond to hypermultiplets of charges
\begin{equation}
 \begin{aligned}
  & \tfrac{1}{2}\big(\delta_1+ \delta_3\pm \delta_0\big), && \pm\tfrac{1}{2}\big[(\delta_1-\delta_3)\mp \delta_0\big],\\
& \delta_{2}+\tfrac{1}{2}\big(\delta_1+ \delta_3\pm \delta_0\big), && \delta_{4}+\delta_{2}+\tfrac{1}{2}\big(\delta_1+ \delta_3\pm \delta_0\big),
 \end{aligned}
\end{equation}
which, together with their PCT conjugates, produce exactly the weights of the $\tfrac{1}{2}(\mathbf{2},\mathbf{8})$.
Again, this is obvious, since the adjoint of $SO(10)$ breaks under the subgroup $SO(8)$ into the adjoint plus two
$\mathbf{8}_v$'s.

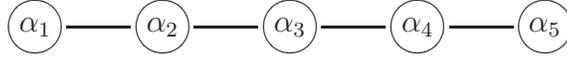
\begin{figure}
\begin{center}
 \begin{equation*}
  \xymatrix{
  *++[o][F-]{{\a}_{1}}\ar@{-}[r]&*++[o][F-]{{\a}_{2}}\ar@{-}[r]&*++[o][F-]{{\a}_3}\ar@{-}[r] & *++[o][F-]{{\a}_4} \ar@{-}[r] & *++[o][F-]{{\a}_5}}
 \end{equation*}
 \end{center}
 \caption{The Dynkin diagram of $SU(6)$.}\label{620dyn}
 \end{figure}

\subsection{$SU(6)$ coupled to $\tfrac{1}{2}\,\mathbf{20}$}\label{sec:620}

\subsubsection{Inducing the quiver} 

With reference to the Dynkin graph in figure \ref{620dyn}, the two Higgs decouplings
\begin{equation}
 \begin{aligned}
&   \alpha_a(\langle\Phi\rangle)= i\, t\, \delta_{a\,2}+O(1)\\
&   \alpha_a(\langle\Phi\rangle)= i\, t\, (\delta_{a\,2}+\delta_{a\,4})+O(1)
 \end{aligned}
\end{equation}
lead to the $SU(2)\times SO(6)$ and the $SU(2)^3$ models, respectively. Comparing the two processes, we see that the full $SU(6)$ quiver is obtained by a symmetrization of the construction of the $SU(2)\times SU(6)$ one with respect to the $\Z_2$ automorphism of the Dynkin graph in fig.\,\ref{620dyn} which interchanges $\alpha_2\leftrightarrow \alpha_4$. In particular, there are no arrows connecting the two Kronecker subquivers which are introduced in the two steps of the inverse Higgs procedure. Hence the pair $(Q,\cw)$ for the $SU(6)$ $\tfrac{1}{2}\,\mathbf{20}$ model is
\begin{equation}\label{quiv620}
\begin{gathered}
\xymatrix{
&*++[o][F-]{1} \ar[ld]_{\phi} \ar[rr]^{H_1} && *++[o][F-]{2}\ar[dl]_{H_2}\ar[dr]^{\widetilde{\psi}}&\\
*++[o][F-]{\kappa} \ar@<0.5ex>@{->}[ddd]^{A}\ar@<-0.5ex>@{->}[ddd]_{B} \ar[ddr]^{\phi^{\prime}}& & *++[o][F-]{3}\ar[lu]_{H_3} \ar[ll]_{\psi \qquad} \ar[rr]^{\quad\widetilde{\phi}}&&*++[o][F-]{\widetilde{\kappa}}\ar@<0.5ex>@{->}[ddd]^{\widetilde{A}}\ar@<-0.5ex>@{->}[ddd]_{\widetilde{B}}
\ar[dddll]_{\widetilde{\phi}^{\prime}}\\
& &  &&\\
&*++[o][F-]{5}\ar[uuu]_{V_1}\ar[dr]^{h_3} & & *++[o][F-]{6}\ar[ll]_{h_1\qquad}\ar[uuu]_{V_2}&\\
*++[o][F-]{\omega}  \ar[rr]^{\psi^{\prime}}& & *++[o][F-]{4}\ar[ur]^{h_2}\ar[uuu]_{V_3} &&*++[o][F-]{\widetilde{\omega}}\ar[ul]_{\widetilde{\psi}^{\prime}}\\
}
\end{gathered}\qquad
\begin{aligned}
&\cw = H_1 H_3 H_2 + h_3 h_1 h_2 +\\
&+ A \psi V_3 \psi^{\prime} + B \psi H_2 V_2 h_2 \psi^{\prime} +\\
& + \phi V_1 \phi^{\prime} + \psi V_3 h_3 \phi^{\prime} +\\
& + \phi H_3 V_3 \psi^{\prime} B+\\
&+ \widetilde{A} \widetilde{\psi} V_2 \widetilde{\psi}^{\prime} + \widetilde{B} \widetilde{\psi} H_1 V_1 h_1 \widetilde{\psi}^{\prime} +\\
& + \widetilde{\phi} V_3 \widetilde{\phi}^{\prime} + \widetilde{\psi} V_2 h_2 \widetilde{\phi}^{\prime} +\\
& + \widetilde{\phi} H_2 V_2 \widetilde{\psi}^{\prime} \widetilde{B}.\end{aligned}
\end{equation}
In principle, $\cw$ may contain further terms of higher order in the 1d Higgs fields (at least order $8$). We consider the computation of the light vector spectrum below as strong evidence that no higher term is needed.
\smallskip

The identifications of the dimension vectors of the $W$--bosons associated to the $A_5$ simple roots (cfr.\! the Dynkin graph in figure \ref{620dyn}) is 
\begin{equation}\label{ideideysu6}
\begin{aligned}
&\alpha_1\rightarrow \d_1\equiv\fd_1 && \alpha_2\to \d_2\equiv\fd_{K_{{0}}} && \a_3\to \d_3\equiv \fd_3 \\
&\a_4\to \d_4\equiv \fd_{K_{\widetilde{0}}} &&\a_5\to \d_5\equiv \fd_2.
\end{aligned}
\end{equation}

\subsubsection{Light BPS spectrum at weak coupling}

Going trough the same steps as for the previous models, we see that the light perturbative category $\mathscr{L}(Q,\cw)$ is equivalent to $\mathsf{rep}(Q^\prime,\cw^\prime)$ where\footnote{\ In passing from $\cw$ to $\cw^\prime$ we set $V_i, B, \widetilde{B}$ equal to $1$ and redefine $h_1,h_3,\widetilde{\psi}^\prime,\widetilde{A}$ by a change of sign.}
\begin{equation}\label{geom620}
Q^{\,\prime}\equiv
\begin{gathered}
\xymatrix@R=2.0pc@C=3.0pc{
&&&&&1 \ar@/_0.5pc/[dll]_{h_3} \ar@/_0.5pc/[dd]_{H_1}\\
&\ar@(ul,dl)[]_{A} 0 \ar@/_0.7pc/[rr]_{\psi^{\prime}}  && 3 \ar@/_0.7pc/[ll]_{\psi} \ar@/_0.5pc/[urr]_{H_3} \ar@/_0.5pc/[drr]_{h_2}&&\\
&&&&&2 \ar@/_0.5pc/[ull]_{H_2} \ar@/_0.5pc/[uu]_{h_1} \ar@/_0.7pc/_{\widetilde{\psi}}[rr]&&\ar@(ur,dr)[]^{\widetilde{A}} \widetilde{0}\ar@/_0.7pc/_{\widetilde{\psi}^\prime}[ll]
}
\end{gathered}
\end{equation}
\begin{equation}\label{geom620+}
 \cw^\prime= H_1H_3H_2+h_3h_1h_2+ A\psi\psi^\prime+\psi(H_2h_2+h_3H_3)\psi^\prime+ \widetilde{A}\widetilde{\psi}\widetilde{\psi}^\prime+\widetilde{\psi}(H_1h_1+h_2H_2)\widetilde{\psi}^\prime.
\end{equation}

In appendix \ref{app:aa} it is verified that for all $X\in \mathsf{rep}(Q^\prime,\cw^\prime)$ the following map
\begin{multline}\label{whatell}
 \ell\colon (X_0, X_{\widetilde{0}}, X_1,X_2,X_3)\longmapsto\\
\longmapsto \big(-AX_0, -\widetilde{A}X_{\widetilde{0}},(h_1H_1+H_3h_3)X_1, (H_1h_1+h_2H_2)X_2,(H_2h_2+h_3H_3)X_3\big),
\end{multline}
is an element of $\mathrm{End}\,X$. Hence, for all light stable $X$, $\ell$ is a complex number $\lambda$. 
\smallskip

The BPS vector--multiplets which remain light in the decoupling limit $g_\mathrm{YM}\rightarrow 0$ (that is, the BPS vector--multiplets which are stable at weak coupling and have \textit{zero} magnetic weights) are required, by physical consistency, to form precisely one copy of the adjoint representation of the gauge group $SU(6)$. In facts, universality of the gauge interactions implies the somehow stronger constraint that the vector sector of $
\mathscr{L}(Q,\cw)$ should be identical to the light category of pure SYM with the same gauge group \cite{cattoy}.

The proof is less elementary than for the two previous models, but the result will be stronger in the sense that not only it shows that the light 
BPS vectors have the right quantum numbers to form the adjoint of $G$, but actually it gives a map to the pure SYM representations which holds chamber by chamber (in the weak coupling regime).

\subsubsection{Universality of the YM sector}\label{sec:univeraality}

It follows from eqn.\eqref{whatell} that
all stable\footnote{\ Here `stable' means stable in \textit{some} weakly coupled chamber.} representations $X\in \mathsf{rep}(Q^\prime,\cw^\prime)$  have
\begin{gather}\label{thhhtterd0}
H_1h_1+h_2H_2=
H_2h_2+h_3H_3=H_3h_3+h_1H_1=\lambda\in \C.
\end{gather}
As in \S\S.\,\ref{light26},\,\ref{light28}, $\lambda$ parameterizes the $\mathbb{P}^1$--family of stable representations associated to a light BPS vector--multiplet, while light BPS hypermultiplets correspond to stable representations which exist at $\lambda=0$ and are \emph{rigid}, that is, have no continuous deformation with $\lambda\neq 0$. Therefore, the light vector--multiplets correspond to stable representations in $\mathsf{rep}(Q^\prime,\cw^\prime)$  which exist for \emph{generic} $\lambda$. 
The universality of the YM sector then implies that for $\lambda\neq 0$ we should get the same representations as in pure SYM at the same (non--zero) value of $\lambda$. In \cite{cattoy} it was found that the set of allowed stable light representations for $\cn=2$ SYM with gauge group $G$
is given, at fixed $\lambda$, by the bricks of the preprojective algebra $\cp(G)$, whose dimension vectors are given by the positive roots of $G$, thus guaranteeing that they form the adjoint representation of the gauge group (see \cite{cattoy} for details). 

In conclusion, universality of the gauge sector requires the bricks of $\mathsf{rep}(Q^\prime,\cw^\prime)$ for the pair \eqref{geom620}\eqref{geom620+} having $\lambda\neq 0$ to be equivalent  to the bricks of the preprojective algebra $\cp(A_5)$.
Once this equivalemce is established, we are guaranteed, in particular, that the BPS vector--multiplets which are stable and light at $g_\mathrm{YM}\sim 0$ make exactly one copy of the adjoint of $SU(6)$ \cite{cattoy}.\medskip

In appendix \ref{app:proofs} it is shown that the arrows of a brick $X\in\mathsf{rep}(Q^\prime,\cw^\prime)$ satisfy the following equations
\begin{equation}\label{thhhtterd1}
h_iH_i(\lambda-h_iH_i)=0 \qquad \text{\begin{small}(not summed over $i$)\end{small}}
\end{equation}
where the index $i=1,2,3$ is defined mod 3. Then, at $\lambda\neq 0$, we may define
\begin{equation}
P_i=\lambda^{-1}\,h_iH_i,\qquad Q_i=\lambda^{-1}\,H_{i-1}h_{i-1},
\end{equation}
and eqn.\eqref{thhhtterd0}\eqref{thhhtterd1} imply that $P_i$ and $Q_i$ are complementary idempotents
\begin{equation}
\begin{aligned}
&P_i^2=P_i, &&Q_i^2=Q_i,\\
&P_iQ_i=Q_iP_i=0,\\
&P_i+Q_i=\mathrm{id}_{X_i}.
\end{aligned}
\end{equation}
Hence
\begin{equation}
X_i=P_iX_i\oplus Q_iX_i,
\end{equation}
and we have the pairs of inverse isomorphims
\begin{equation}\label{isosio}
\xymatrix{P_iX_i\ar@<0.7ex>[rr]^{H_iP_i}&& Q_{i+1}X_{i+1}.\ar@<0.7ex>[ll]^{h_iQ_{i+1}}}
\end{equation}

Identifying isomorphic subspaces trough the maps \eqref{isosio}, a brick $X\in\mathsf{rep}(Q^\prime,\cw^\prime)$ may be seen as a representation $X^\flat$ of the following quiver
\begin{equation}\label{quiquoqua0}
\begin{gathered}
\xymatrix{ P_3X_3\ar@/_1.2pc/[dd]_{\psi P_3}\\
\\
 0\ar@/_1.2pc/[uu]_{P_3\psi^\prime}\ar@/_1.2pc/[rr]_{Q_3\psi^\prime} && Q_3X_3\ar@/_1.2pc/[ll]_{\psi Q_3}
\ar@/_1.2pc/[rr]_{\lambda^{-1}\widetilde{\psi}P_2h_2}
&& \widetilde{0}\ar@/_1.2pc/[ll]_{Q_3H_2\widetilde{\psi}^\prime}
\ar@/_1.2pc/[rr]_{Q_2\widetilde{\psi}^\prime}&&
Q_2X_2\ar@/_1.2pc/[ll]_{\widetilde{\psi}Q_2}},
\end{gathered}
\end{equation}
which is the \emph{double}\footnote{\ Given a graph $\Gamma$, its \textit{double}, $\overline{\Gamma}$, is the quiver obtained by replacing each link of $\Gamma$ by a pair of opposite arrows $\leftrightarrows$.} $\overline{A_5}$ of the $A_5$ Dynking graph.
Moreover, the following relations hold (see appendix \ref{app:proofs} for a proof)
\begin{align}\label{relRELxxy5}
&\psi P_3\cdot P_3\psi^\prime+
\psi Q_3\cdot Q_3\psi^\prime=0\\
\label{relRELxxy52}&P_3\psi^\prime\cdot \psi P_3=Q_2\widetilde{\psi}^\prime\cdot \widetilde{\psi}Q_2=0\\
\label{relRELxxy53}&\widetilde{\psi}Q_2\cdot Q_2\widetilde{\psi}^\prime+\lambda^{-1}\,\widetilde{\psi}P_2h_2\cdot Q_3H_2\widetilde{\psi}^\prime=0\\
&Q_3\psi^\prime\cdot \psi Q_3+Q_3H_2\widetilde{\psi}^\prime\cdot (\lambda^{-1}\,\widetilde{\psi}P_2h_2)=0 \label{relRELxxy10}
\end{align}

The path algebra of the double quiver $\overline{A_5}$ with the relations 
\eqref{relRELxxy5}--\eqref{relRELxxy10} defines the preprojective algebra  $\cp(A_5)$ (see \cite{cattoy} and references therein). \smallskip

Hence we get a map 
\begin{equation}X\mapsto X^\flat\in \mathsf{mod}\,\cp(A_5)\end{equation}
which is well defined on the subcategory (object class)
$\mathscr{B}_{\lambda\neq 0}\subset \mathsf{rep}(Q^\prime,\cw^\prime)$ of representations satisfying the extra relations \eqref{thhhtterd0} and \eqref{thhhtterd1} with $\lambda\neq 0$. For $X\in\mathscr{B}_{\lambda\neq 0}$ we have
\begin{equation}
\text{End }X\simeq \text{End }X^\flat,
\end{equation}
and hence \begin{equation}X\in\mathsf{rep}(Q^\prime,\cw^\prime)\ \text{is a brick with }\lambda\neq 0\quad \Longleftrightarrow\quad X^\flat\in \mathsf{mod}\,\cp(A_5)\ \text{is a brick.}\end{equation}
 An indecomposable representation $X^\flat\in\mathsf{mod}\,\cp(A_5)$ with\footnote{\ Here $e_a$ stands for the dimension vector which is $1$ at the $a$--th node of the quiver of $\cp(G)$ (namely the \textit{double} quiver of the Dynkin graph $G$)  and zero elsewhere. The correspondence $e_a\leftrightarrow \alpha_a$ identifies the dimension lattice of $\cp(G)$ with the root lattice of $G$. } $\dim X^\flat=\sum_a n_a\,e_a$ is a brick (necessarily rigid) if and only if $\dim X^\flat=\sum_a n_a\, \alpha_a$ is a positive root of $A_5$ \cite{cattoy}.

To complete the argument,
it remains to notice that, under the map $X\mapsto X^\flat$, 
the representations of the affine subquivers $\mathbf{Kr}_s$ and $\widehat{A}(3,1)_i$
 which correspond to the simple--root $W$--bosons introduced trough the recursive inverse Higgs procedure --- whose dimension vectors $\delta_a$ are listed in eqn.\eqref{ideideysu6} --- get mapped to the simple representations of $\cp(G)$ with dimension vectors $e_a$
 \begin{equation}
 X\in\mathscr{B}_{\lambda\neq 0}\ \text{and }\dim X=\delta_a\quad\Longrightarrow\quad \dim X^\flat=e_a.
 \end{equation}

Therefore, if $X\in\mathsf{rep}(Q^\prime,\cw^\prime)$ is a brick with $\lambda\neq 0$, we have
\begin{equation}\label{rrrttq}
\dim X=\sum_a n_a\,\delta_a\qquad\text{with }\sum_an_a\,\a_a\ \text{a positive root of }SU(6).
\end{equation}
Moreover, stability  of $X\in\mathscr{B}_{\lambda\neq0}$ with respect to a given induced central charge with $Z(\delta_a)=Z_a$
is equivalent to the stability of $X^\flat\in\mathsf{mod}\,\cp(A_5)$ with respect to the central charge $Z(e_a)=Z_a$, that is, to the stability of the corresponding pure $SU(6)$ SYM representation with respect to the natural central charge.
Therefore, the light BPS vectors form precisely a copy of the adjoint of $SU(6)$, uniqueness following from uniqueness for the pure SYM case (see \cite{cattoy}).

In general, there are many bricks of $\cp(A_5)$ with dimension vector equal to a given positive root of $A_5$. However, in each (weakly coupled) chamber precisely one representation per root is stable \cite{ACCERV2,cattoy}. Indeed, in eqn.\eqref{rrrttq} one has $n_a=0,1$ and hence all arrows in a brick are either isomorphisms or zero. This implies that for each pair of opposite arrows in the quiver \eqref{quiquoqua0} one arrow vanishes. Choosing (say)
\begin{equation}\label{wwwrtzq}
 \arg Z(\fd_1)=\arg Z(\fd_2)=\arg Z(\fd_3)=\pi/2,\qquad \arg Z(\fd_{K_0})=\arg Z(\fd_{K_{\widetilde{0}}})=0,
\end{equation}
the vanishing arrows are $\widetilde{\psi}^\prime$ and $\psi^\prime$. 

\subsubsection{Light BPS hypermultiplets}\label{sec:lighthyper}

The hypermultiplets correspond to stable representations of $\mathsf{rep}(Q^\prime,\cw^\prime)$ with $\lambda=0$ which are not in the projective closure of the family of a vector--multiplet. We may assume $X_{0},X_{\widetilde{0}}\neq 0$, since otherwise we reduce to representations of subquivers which were already studied in \S\S.\,\ref{light26},\,\ref{light28}. 

The analysis is much more involved that in the $\lambda\neq 0$ case. Instead of analyzing the general brick of $(Q^\prime,\cw^\prime)$ subjected to the extra constraints
\begin{equation}
\label{thhhtterd00}
H_1h_1+h_2H_2=
H_2h_2+h_3H_3=H_3h_3+h_1H_1=0,
\end{equation}
we consider some \emph{special} classes of such representations and show that --- restricting to these special classes ---
there are no light hypermultiplets with exotic quantum numbers. We expect the result to be true in full generality, although we have no complete proof.

The special classes we focus on have the  additional property that one of the four arrows
$\psi, \psi^\prime, \widetilde{\psi}^\prime, \widetilde{\psi}$ vanishes. If, say, $\widetilde{\psi}^\prime=0$, we have a subrepresentation 
\begin{equation}\label{subspacesY34}0\rightarrow \big(0,\mathrm{Im}(\widetilde{\psi}h_2\psi^\prime\psi),\text{Im}(H_3\psi^\prime\psi),
\text{Im}(h_2\psi^\prime\psi),\text{Im}(\psi^\prime\psi)\big)\rightarrow (X_0,X_{\widetilde{0}},X_1,X_2,X_3)\end{equation}
which, in the chamber \eqref{wwwrtzq}, is destablizing unless it vanishes. Then $\psi^\prime\psi=0$, and from the relations $\partial\cw^\prime=0$ we get
\begin{equation}\label{thhhtterd001}
 H_3H_2=H_1H_3=H_2H_1=h_2h_3=h_3h_1=h_1h_2=0.
\end{equation}
It follows from eqn.\eqref{thhhtterd00}\eqref{thhhtterd001}
that
\begin{equation}
 (X_0,X_{\widetilde{0}},X_1,X_2,X_3)\mapsto (0,0,h_1H_1X_1,0,0,0)
\end{equation}
is a nilpotent element of $\mathrm{End}\,X$ hence zero (since $X$ is a brick). Then \begin{equation}\label{ppppq321}h_1H_1=H_3h_3=0.\end{equation}

As a consequence of eqns.\eqref{thhhtterd001}\eqref{ppppq321}, a stable representation has $h_1=H_3=0$. Indeed, if $\text{Im }h_1\neq 0$ (resp.\! $\text{Im }H_3\neq 0$),  the subrepresentation $Y$ with $Y_1=\text{Im }h_1$ (resp.\! $Y_1=\text{Im }H_3$) and zero elsewhere would be destabilizing (by assumption, $X_0,X_{\widetilde{0}}\neq 0$). From eqn.\eqref{thhhtterd00} we get
\begin{equation}
 h_2H_2=H_2h_2=0.
\end{equation}
Choosing 
\begin{equation}|Z_0|\gg |Z_2|\ggg |Z_{\widetilde{0}}|, |Z_1|, |Z_2|,\end{equation}
the subrepresentation
\begin{equation}
 0\rightarrow \Big(0, \text{Im}(\psi h_2+\psi H_1), 0,\text{Im }h_2+\text{Im }H_1,0\Big)\rightarrow (X_0,X_{\widetilde{0}},X_1, X_2, X_3)
\end{equation}
is destabilizing unless $\text{Im }h_2+\text{Im }H_1=0$. Hence $H_1=h_2=0$. 

\begin{figure}
 \begin{align*}
  &\xymatrix{ & 0\ar@{-}[d] &\\ 1\ar@{-}[r]& 1\ar@{-}[r] & 1\ar@{-}[r]&1} 
&&\xymatrix{ & 1\ar@{-}[d] &\\ 1\ar@{-}[r]& 2\ar@{-}[r] & 1\ar@{-}[r]\ar@{-}[r]&1}\\
&\phantom{---}\\
&\xymatrix{ & 1\ar@{-}[d] &\\  1\ar@{-}[r] & 1\ar@{-}[r]&1\ar@{-}[r]&1}
&&\xymatrix{ & 1\ar@{-}[d] &\\ 1 \ar@{-}[r] & 2\ar@{-}[r]&2\ar@{-}[r]&1}
 \end{align*}
\caption{\label{fiurD57}The four positive roots of $D_5$ with non--zero $X_0,X_{\widetilde{0}}$. }
\end{figure}
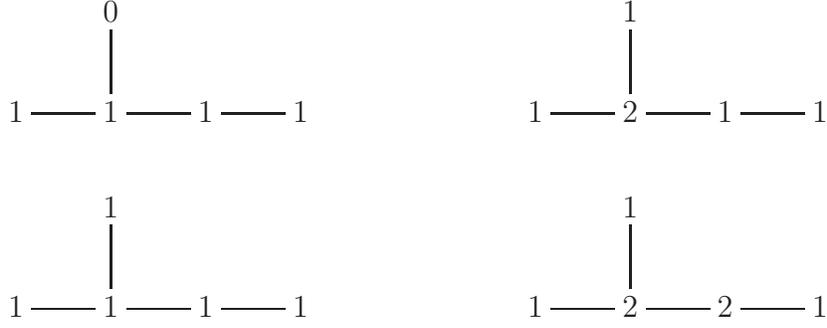

In conclusion --- under the simplifying assumption $\widetilde{\psi}^\prime=0$ --- a stable representation $X$ with $X_0,X_{\widetilde{0}}\neq 0$ is effectively a representation of the quiver 
\begin{equation}\label{rrrrvv}
\begin{gathered}\xymatrix{&& 1\ar[dl]_{h_3}\\
0 & 3\ar[l]_\psi  && 2\ar@/_0.7pc/[ll]_{H_2}\ar[r]^{\widetilde{\psi}}& \widetilde{0}}\end{gathered}
\end{equation}
which is a $D_5$ Dynkin quiver, whose stable representations are given by the positive roots. Hence the states we are looking for have dimension vectors equal to the positive roots of $D_5$ whose support contains both $0$ and $\widetilde{0}$. There are four such roots represented in figure \ref{fiurD57}.

The \emph{physical} charge vectors associated to these four representation are (with reference to the Dynkin graph in figure \ref{620dyn})
\begin{equation}
 \begin{aligned}
  &\delta_2+\delta_3+\delta_4 &&\delta_2+\delta_3+\delta_4+\delta_5\\
&  \tfrac{1}{2}\delta_1+\delta_2+\tfrac{1}{2}\delta_3+\delta_4+\tfrac{1}{2}\delta_5,
&&\tfrac{1}{2}\delta_1+\delta_2+\tfrac{3}{2}\delta_3+\delta_4+\tfrac{1}{2}\delta_5.
 \end{aligned}
\end{equation}
The first two are in the closure of the $\mathbb{P}^1$--families and are part of the vector--multiplets. The last two correspond to new stable hypermultiplets. Indeed, adding to these two the eight hyper representations with vanishing $X_0$ or $X_{\widetilde{0}}$ (whose charge vectors are given by eqn.\eqref{yyyyxxxbbp}) we get precisely the weights of the $\tfrac{1}{2}\,\mathbf{20}$ (in our basis) namely,
\begin{equation}
 \begin{aligned}
&  \tfrac{1}{2}\delta_1+\delta_2+\tfrac{3}{2}\delta_3+\delta_4+\tfrac{1}{2}\delta_5, 
&&\quad\tfrac{1}{2}\delta_1+\delta_2+\tfrac{1}{2}\delta_3+\delta_4+\tfrac{1}{2}\delta_5,\\  
&\tfrac{1}{2}\delta_1+\tfrac{1}{2}\delta_3+\delta_4+\tfrac{1}{2}\delta_5,
&&\quad\tfrac{1}{2}\delta_1+\delta_2+\tfrac{1}{2}\delta_3+\tfrac{1}{2}\delta_5,\\
&\tfrac{1}{2}\delta_1+\tfrac{1}{2}\delta_3+\tfrac{1}{2}\delta_5,
&&\quad\tfrac{1}{2}\delta_1+\delta_2+\tfrac{1}{2}\delta_3-\tfrac{1}{2}\delta_5,\\ 
&-\tfrac{1}{2}\delta_1+\tfrac{1}{2}\delta_3+\delta_4+\tfrac{1}{2}\delta_5,
&&\quad\tfrac{1}{2}\delta_1-\tfrac{1}{2}\delta_3+\tfrac{1}{2}\delta_5,\\ 
&\tfrac{1}{2}\delta_1+\tfrac{1}{2}\delta_3-\tfrac{1}{2}\delta_5,
&&\quad-\tfrac{1}{2}\delta_1+\tfrac{1}{2}\delta_3+\tfrac{1}{2}\delta_5. 
 \end{aligned}
\end{equation}

In particular, no exotic quantum number appear. Although the proof refers to some special classes, the analogy with pure SYM suggests that there are chambers in which all stable light states do correspond to representation of such a special class, and hence that the result is true in general, as required by physical consistency.

\subsection{$SO(12)$ coupled to $\tfrac{1}{2}\,\textbf{32}$ and $E_7$ coupled to $\tfrac{1}{2}\,\textbf{56}$}\label{sec:756}

For the last two models in figure  \ref{mapofinduction}, $SO(12)$ coupled to $\tfrac{1}{2}\,\textbf{32}$ and $E_7$ coupled to $\tfrac{1}{2}\,\textbf{56}$, the induction of the quiver by the inverse Higgs procedure is quite simple since at each step the new Kronecker quiver is connected only to a Kronecker subquiver added in a previous step of the procedure. This corresponds to the group--theoretical fact that these models have alternate Higgs decoupling schemes, with positive linear forms $\lambda$, which reduce the theory, respectively, to pure $SU(3)$ SYM and pure $SU(4)$ SYM.
\smallskip

\begin{figure}
\begin{equation*}
\begin{gathered}
\xymatrix{
&&&&&*++[o][F-]{1} \ar[ld]_{\phi} \ar[rr]^{H_1} && *++[o][F-]{2}\ar[dl]_{H_2}\ar[dr]^{\widetilde{\psi}}&\\
*++[o][F-]{\tau_{-2}}\ar@<0.5ex>@{->}[ddd]^{A_{-2}}\ar@<-0.5ex>@{->}[ddd]_{B_{-2}}&& *++[o][F-]{\omega_{-1}}\ar[ll]_{\psi_{-2}} \ar[rr]^{\psi_{-1}^{\prime}}&&*++[o][F-]{\tau_0} \ar@<0.5ex>@{->}[ddd]^{A_0}\ar@<-0.5ex>@{->}[ddd]_{B_0} \ar[ddr]^{\phi^{\prime}}& & *++[o][F-]{3}\ar[lu]_{H_3} \ar[ll]_{\psi_0 \qquad} \ar[rr]^{\quad\widetilde{\phi}}&&*++[o][F-]{\widetilde{\tau}}\ar@<0.5ex>@{->}[ddd]^{\widetilde{A}}\ar@<-0.5ex>@{->}[ddd]_{\widetilde{B}}\ar[dddll]_{\widetilde{\phi}^{\prime}}\\
&&&&& &  &&\\
&& &&&*++[o][F-]{5}\ar[uuu]_{V_1}\ar[dr]^{h_3} & & *++[o][F-]{6}\ar[ll]_{h_1\qquad}\ar[uuu]_{V_2}&\\
*++[o][F-]{\omega_{-2}}\ar[rr]^{\psi_{-2}^{\prime}}&&*++[o][F-]{\tau_{-1}} \ar@<0.5ex>@{->}[uuu]^{A_{-1}}\ar@<-0.5ex>@{->}[uuu]_{B_{-1}}&&*++[o][F-]{\omega_0}\ar[ll]_{\psi_{-1}}  \ar[rr]^{\psi_0^{\prime}}& & *++[o][F-]{4}\ar[ur]^{h_2}\ar[uuu]_{V_3} &&*++[o][F-]{\widetilde{\omega}}\ar[ul]_{\widetilde{\psi}^{\prime}}\\
}
\end{gathered}
\end{equation*}
\caption{\label{last2quivers}The quiver of $E_7$ coupled to $\tfrac{1}{2}\,\textbf{56}$. Omitting the Kronecker subquiver over the nodes $\tau_{-2},\omega_{-2}$ we get the quiver for
$SO(12)$ coupled to $\tfrac{1}{2}\,\textbf{32}$.}
\end{figure}
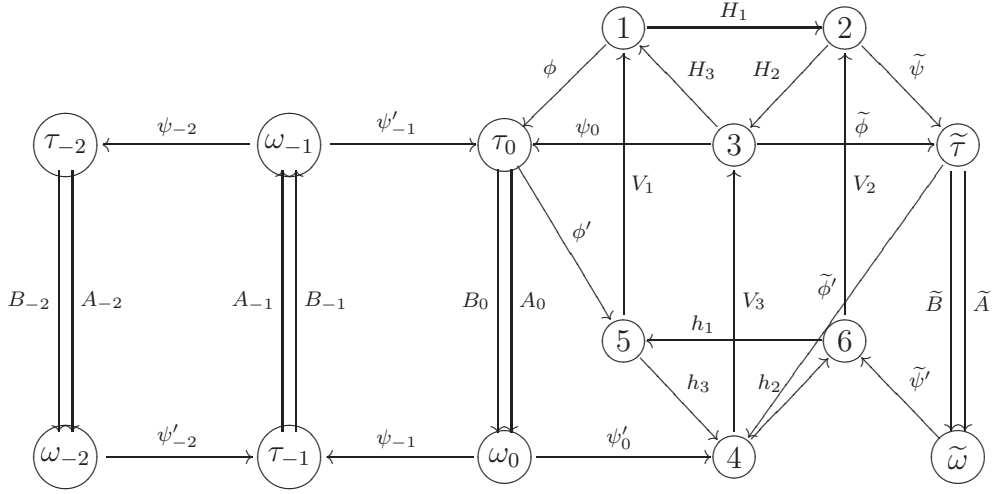

The quiver of $E_7$ coupled to $\tfrac{1}{2}\,\textbf{56}$ is presented in figure \ref{last2quivers}; omitting the two nodes $\tau_{-2},\omega_{-2}$, which corresponds to the Higgs decoupling $E_7\rightarrow SO(12)$, we get the one for
$SO(12)$ coupled to $\tfrac{1}{2}\,\textbf{32}$.

The superpotentials $\cw$ are also determined by the consistency of the various Higgs decoupling schemes. For the $E_7$ case $\cw$ is
\begin{equation}\label{wwwwthis}\begin{split}
 \cw_{E_7}&=\cw_{SU(6)}+\cw_{SYM}=\\
&=\cw_{SU(6)}+A_0\psi^\prime_{-1}B_{-1}\psi_{-1}-B_0\psi^\prime_{-1}A_{-1}\psi_{-1}+\\
&\qquad+A_{-1}\psi^\prime_{-2}B_{-2}\psi_{-2}-B_{-1}\psi^\prime_{-2}A_{-2}\psi_{-2},
\end{split}\end{equation}
where $\cw_{SU(6)}$ is the one in eqn.\eqref{quiv620} while $\cw_{SYM}$ is the one for pure $SU(4)$ SYM \cite{ACCERV2,cattoy}.
$\cw_{SO(12)}$ is  obtained by omitting the last line in eqn.\eqref{wwwwthis}. Again, in principle $\cw$ may contain additional terms of order at least 10 in the arrows, which would be relevant only for high dimension representations.

 \begin{figure}
\begin{center}
 \begin{equation*}
  \xymatrix{&& *++[o][F-]{{\a}_{2}}\ar@{-}[d]\\
  *++[o][F-]{{\a}_{1}}\ar@{-}[r]&*++[o][F-]{{\a}_{3}}\ar@{-}[r]&*++[o][F-]{{\a}_4}\ar@{-}[r] & *++[o][F-]{{\a}_5} \ar@{-}[r] & *++[o][F-]{{\a}_6} \ar@{-}[r] & *++[o][F-]{{\a}_7}}
 \end{equation*}
 \end{center}
 \caption{The Dynkin diagram of $E_7$.}\label{756dyn}
 \end{figure}
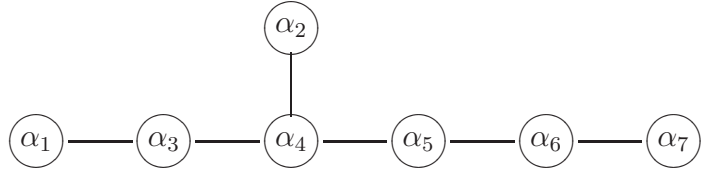

\smallskip

The identifications of the dimension vectors of the $W$--bosons associated to the $E_7$ simple roots (cfr.\! the Dynkin graph in figure \ref{756dyn}) is 
\begin{equation}\label{ideidey}
\begin{aligned}
&\alpha_1\rightarrow \d_1\equiv\fd_{K_{-2}} &&\alpha_2\rightarrow \d_2\equiv\fd_1 && \alpha_3\to \d_3\equiv\fd_{K_{-1}} && \alpha_4\to \d_4\equiv\fd_{K_0}\\
&\a_5\to \d_5\equiv \fd_3 &&\a_6\to \d_6\equiv \fd_{K_{\widehat{0}}}
&&\a_7\to \d_7\equiv \fd_2.
\end{aligned}
\end{equation}

\subsubsection{The light sector}
Let us consider first $SO(12)$ SYM coupled to $\tfrac{1}{2}\,\mathbf{32}$.
Assuming consistency with the perturbative Higgs effect, as long as we are interested only in the light category $\mathscr{L}(Q,\cw)$ (which contains the BPS states at weak coupling with zero magnetic charges \cite{cattoy}), we may set $B_{-1}, B_0, \widetilde{B}$ and the $V_i$'s to $1$, ending up with representations in the category $\mathsf{rep}(Q^\prime,\cw^\prime)$ where
\begin{equation}\label{geomeee}
\begin{gathered}
\xymatrix@R=2.0pc@C=3.0pc{
&&&&&1 \ar@/_0.5pc/[dll]_{h_3} \ar@/_0.5pc/[dd]_{H_1}\\
&\ar@(ul,dl)[]_{A_{-1}} -1 \ar@/_0.7pc/[r]_{\psi_{-1}^{\prime}} &\ar@/_0.7pc/[l]_{\psi_{-1}} \ar@(dr,dl)[]^{A_0} 0 \ar@/_0.7pc/[r]_{\psi_0^{\prime}} & 3 \ar@/_0.7pc/[l]_{\psi_0} \ar@/_0.5pc/[urr]_{H_3} \ar@/_0.5pc/[drr]_{h_2}&&\\
&&&&&2 \ar@/_0.5pc/[ull]_{H_2} \ar@/_0.5pc/[uu]_{h_1}\ar@/_0.7pc/_{\widetilde{\psi}}[rr]&&\ar@(ur,dr)[]^{\widetilde{A}} \widetilde{0}\ar@/_0.7pc/_{\widetilde{\psi}^\prime}[ll]
}
\end{gathered}
\end{equation}
and
\begin{multline}
 \cw^\prime= H_1H_3H_2+h_3h_1h_2+\psi_0(H_2h_2-h_3H_3)\psi_0^\prime+ \\ +\widetilde{\psi}(H_1h_1-h_2H_2)\widetilde{\psi}^\prime
+A_0\psi_0\psi_0^\prime+A_0\psi^\prime_{-1}\psi_{-1}-A_{-1}\psi_{-1}\psi^\prime_{-1}
+\widetilde{A}\widetilde{\psi}\widetilde{\psi}^\prime.
\end{multline}
 The main difference with respect to the previous case, is that the relations no longer imply $(\psi_0^\prime\psi_0)^2=(\widetilde{\psi}^\prime\widetilde{\psi})^2=0$ but rather the weaker conditions
 \begin{equation}(\psi_0^\prime\psi_0)^3=(\widetilde{\psi}^\prime\widetilde{\psi})^2
 =0.\label{ReLer}\end{equation} This fact makes the analysis of the brick representations much harder than in the previous model. Again, one shows that the following linear map
\begin{multline}
 \ell\colon (X_{-1}, X_0, X_{\widetilde{0}}, X_1, X_2, X_3)\rightarrow \Big(A_{-1}X_{-1}, A_0X_0, -\widetilde{A}X_{\widetilde{0}},\\
 (h_1H_1+H_3h_3)X_1, (H_1h_1-h_2H_2)X_2, \big(h_3H_3-H_2h_2-(\psi^\prime_0\psi_0)^2\big)X_3\Big)
\end{multline}
is an element of $\mathrm{End}\,X$, ande hence a complex number $\lambda$ if $X$ is a brick.

The analysis of the general representations of the Jacobian algebra $\C Q^\prime/(\partial \cw^\prime)$ is rather hard.
However, things simplify for special classes of representations. If, we assume any one of the two conditions
\begin{equation}\label{assumpppt}
(\psi_0^\prime\psi_0)^2=0,\qquad \text{or}\qquad \widetilde{\psi}^\prime\widetilde{\psi}=0,
\end{equation} 
which are only slightly stronger than the corresponding relation \eqref{ReLer}, we reduce to the same analysis as in the preceding sections. By going trough the same steps we see that, at $\lambda\neq 0$, we would end up with a map
\begin{equation}
\begin{aligned}
(\cdot)^\flat\colon  &\mathscr{B}_{\lambda\neq 0}\rightarrow \mathsf{mod}\, \cp(D_6)\\
&X\mapsto X^\flat
\end{aligned}
\end{equation} 
sending the elements of the object class $\mathscr{B}_{\lambda\neq 0}\subset \mathsf{rep}(Q^\prime,\cw^\prime)$ of bricks satisfying $\lambda\neq 0$ \textit{and} \eqref{assumpppt} in representations of the \textit{double} of the $D_6$ Dynkin quiver 
\begin{equation}\label{quiquoqua0}
\overline{D_6}\colon\ \begin{gathered}
\xymatrix{ && P_3X_3\ar@/_1.2pc/[dd]_{\psi_0 P_3}\\
\\
-1\ar@/_1.2pc/[rr]_{\psi_{-1}^\prime} && \ar@/_1.2pc/[ll]_{\psi_{-1}^\prime}0\ar@/_1.2pc/[uu]_{P_3\psi_0^\prime}\ar@/_1.2pc/[rr]_{Q_3\psi_0^\prime} && Q_3X_3\ar@/_1.2pc/[ll]_{\psi_0 Q_3}
\ar@/_1.2pc/[rr]_{\lambda^{-1}\widetilde{\psi}P_2h_2}
&& \widetilde{0}\ar@/_1.2pc/[ll]_{Q_3H_2\widetilde{\psi}^\prime}
\ar@/_1.2pc/[rr]_{Q_2\widetilde{\psi}^\prime}&&
Q_2X_2\ar@/_1.2pc/[ll]_{\widetilde{\psi}Q_2}},
\end{gathered}
\end{equation}
subjected precisely to the relations of the preprojective algebra $\cp(D_6)$. The discussion in \S.\,\ref{sec:univeraality} applies word--for--word, and,  again, the stable light vector--multiplets (in this object class) make a single copy of the adjoint representation of the gauge group $SO(12)$. In particular, this shows that all the $SO(12)$ $W$--boson exists, and  no vector--multiplet with exotic quantum numbers exists in this class.   Physically, by universality of the SYM sector, one expects this to be the full story as far as light vector--multiplets are concerned, and that all the relevant light stable representations actually belong to the class  \eqref{assumpppt}.
\bigskip

For $E_7$ coupled to $\tfrac{1}{2}\,\mathbf{56}$ $Q^\prime$ the light category $\mathscr{L}(Q,\cw)$ --- which contains the BPS states at weak coupling with zero magnetic charges \cite{cattoy} ---  is given by $\mathsf{rep}(Q^\prime,\cw^\prime)$ for the pair $(Q^\prime,\cw^\prime)$
\begin{equation}\label{geomeee}
\begin{gathered}
\xymatrix@R=2.0pc@C=3.0pc{
&&&&&&1 \ar@/_0.5pc/[dll]_{h_3} \ar@/_0.5pc/[dd]_{H_1}\\
&\ar@(ul,dl)[]_{A_{-2}} -2 \ar@/_0.7pc/[r]_{\psi_{-2}^{\prime}} &\ar@/_0.7pc/[l]_{\psi_{-2}} \ar@(dr,dl)[]^{A_{-1}} -1\ar@/_0.7pc/[r]_{\psi_{-1}^{\prime}}&\ar@/_0.7pc/[l]_{\psi_{-1}} \ar@(dr,dl)[]^{A_0} 0 \ar@/_0.7pc/[r]_{\psi_0^{\prime}} & 3 \ar@/_0.7pc/[l]_{\psi_0} \ar@/_0.5pc/[urr]_{H_3} \ar@/_0.5pc/[drr]_{h_2}&&\\
&&&&&&2 \ar@/_0.5pc/[ull]_{H_2} \ar@/_0.5pc/[uu]_{h_1}\ar@/_0.7pc/_{\widetilde{\psi}}[rr]&&\ar@(ur,dr)[]^{\widetilde{A}} \widetilde{0}\ar@/_0.7pc/_{\widetilde{\psi}^\prime}[ll]
}
\end{gathered}
\end{equation}
\begin{multline}
 \cw^\prime= H_1H_3H_2+h_3h_1h_2+\psi_0(H_2h_2+h_3H_3)\psi_0^\prime+ \widetilde{\psi}(H_1h_1+h_2H_2)\widetilde{\psi}^\prime+\\
+A_0\psi_0\psi_0^\prime+A_0\psi^\prime_{-1}\psi_{-1}-A_{-1}\psi_{-1}\psi^\prime_{-1}+A_{-1}\psi^\prime_{-2}\psi_{-2}-A_{-2}\psi_{-2}\psi_{-2}^\prime
+\widetilde{A}\widetilde{\psi}\widetilde{\psi}^\prime.
\end{multline}
The same analysis as for the previous example would lead --- for the $\lambda\neq 0$ bricks of the special class \eqref{assumpppt} --- to representations of the $\cp(E_7)$ preprojective algebra
\begin{equation}
\begin{aligned}
(\cdot)^\flat\colon  &\mathscr{B}_{\lambda\neq 0}\rightarrow \mathsf{mod}\, \cp(E_7)\\
&X\mapsto X^\flat
\end{aligned}
\end{equation} 
where $X^\flat$ is as follows
\begin{equation}\label{quiquoqua007}
\overline{E_7}\colon\ \begin{gathered}
\xymatrix{ &&&& P_3X_3\ar@/_1.2pc/[dd]_{\psi_0 P_3}\\
\\
-2\ar@/_1.2pc/[rr]_{\psi_{-2}^\prime} && \ar@/_1.2pc/[ll]_{\psi_{-2}^\prime}-1\ar@/_1.2pc/[rr]_{\psi_{-1}^\prime} && \ar@/_1.2pc/[ll]_{\psi_{-1}^\prime}0\ar@/_1.2pc/[uu]_{P_3\psi_0^\prime}\ar@/_1.2pc/[rr]_{Q_3\psi_0^\prime} && Q_3X_3\ar@/_1.2pc/[ll]_{\psi_0 Q_3}
\ar@/_1.2pc/[rr]_{\lambda^{-1}\widetilde{\psi}P_2h_2}
&& \widetilde{0}\ar@/_1.2pc/[ll]_{Q_3H_2\widetilde{\psi}^\prime}
\ar@/_1.2pc/[rr]_{Q_2\widetilde{\psi}^\prime}&&
Q_2X_2\ar@/_1.2pc/[ll]_{\widetilde{\psi}Q_2}},
\end{gathered}
\end{equation}
Again, this gives all the $E_7$ $W$--bosons.
\medskip

The analysis of the $\lambda=0$ bricks is even more involved. Although there is no reason to expect that the analysis of the special class of representations would be enough, it is amusing to note that, in the $SO(12)$ case, specializing to a chamber analogous to the one considered in \S.\,\ref{sec:lighthyper}, the stable $\lambda=0$ representations in the special class get mapped into the stable representations of the $E_6$ Dynkin quiver
\begin{equation}\label{rrrrvve6}
\xymatrix{&&& 1\ar[dl]_{h_3}\\
-1 & 0\ar[l]_{\psi_{-1}} & 3\ar[l]_{\psi_0}  && 2\ar@/_0.7pc/[ll]_{H_2}\ar[r]^{\widetilde{\psi}}& \widetilde{0}}
\end{equation}

\begin{figure}
\begin{align*}
  &\xymatrix{ && 1\ar@{-}[d] &\\ 1\ar@{-}[r] &1\ar@{-}[r]& 1\ar@{-}[r] & 1\ar@{-}[r]&1} 
&&\xymatrix{ && 1\ar@{-}[d] &\\ 1\ar@{-}[r]&1\ar@{-}[r]& 2\ar@{-}[r] & 2\ar@{-}[r]\ar@{-}[r]&1}\\
&\phantom{---}\\
&\xymatrix{ && 1\ar@{-}[d] &\\  1\ar@{-}[r] &2\ar@{-}[r]& 2\ar@{-}[r]&2\ar@{-}[r]&1}
&&\xymatrix{ && 2\ar@{-}[d] &\\1\ar@{-}[r] & 2 \ar@{-}[r] & 3\ar@{-}[r]&2\ar@{-}[r]&1}
 \end{align*}
\caption{\label{fiurE6}The 4  roots of $E_6$ with non--zero $X_{-1},X_{\widetilde{0}}$ not in the closure of $W$--bosons. }
\end{figure}
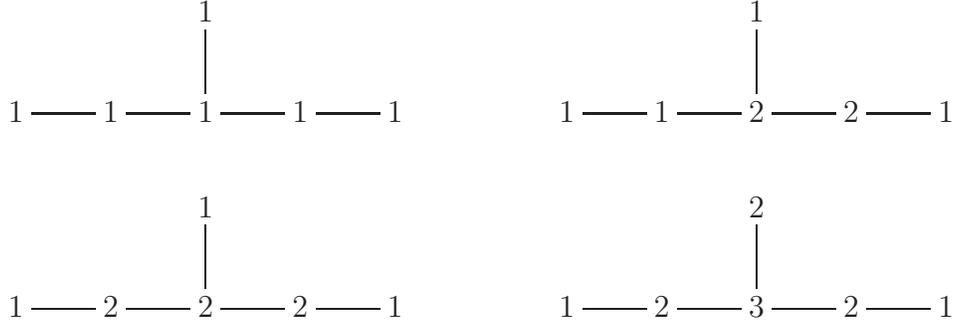 

Stable representations then correspond to positive roots of $E_6$. Those which do not correspond to known representations of subquivers have $X_{-1}, X_0, X_{\widetilde{0}}\neq 0$. There are 8 such roots. 4 are in the closure of $W$--bosons; the remaining 4, represented in fig.\,\ref{fiurE6}, correspond to `new' hypermultiplets which do not come from representations of proper subquivers. Their dimension vectors are
\begin{equation}
 \begin{aligned}
  &\tfrac{1}{2}(\fd_1+\fd_2+\fd_3)+\fd_{K_{\widetilde{0}}}+\fd_{K_{0}}+\fd_{K_{-1}} &&\qquad
\tfrac{3}{2}\fd_3+\tfrac{1}{2}(\fd_1+\fd_2)+\fd_{K_{\widetilde{0}}}+\fd_{K_{0}}+\fd_{K_{-1}}\\
  &\tfrac{3}{2}\fd_3+\tfrac{1}{2}(\fd_1+\fd_2)+\fd_{K_{\widetilde{0}}}+2\,\fd_{K_{0}}+\fd_{K_{-1}} &&\qquad
\tfrac{3}{2}(\fd_1+\fd_3)+\tfrac{1}{2}\,\fd_2+\fd_{K_{\widetilde{0}}}+2\,\fd_{K_{0}}+\fd_{K_{-1}} 
\end{aligned}
\end{equation}
which precisely correspond to the 4 weights of $\tfrac{1}{2}\,\mathbf{32}$ which  are \textit{sincere} (in our basis), that is, all coefficients in their expansions as $\sum_i a_i \fd_i$ are non zero. The other `non--sincere' 12 states have support on subquivers and were analyzed previously. Putting everything together, at weak coupling the \emph{special class} light BPS hypers  make precisely a copy of the $\tfrac{1}{2}\,\mathbf{32}$.
\medskip

The $E_7$ case is more involved, and not all states of the $\tfrac{1}{2}\,\mathbf{56}$ will belong to the special class.

\subsection{Formal models}

The reader has certainly noticed that our
Higgs procedure produces an extra pair $(Q,\cw)$. Indeed, starting from the $E_7$ model and considering the Higgs background
\begin{equation}
\alpha_a(\langle \Phi\rangle)=i\,t\,\delta_{a\,r}+O(1),\qquad t\rightarrow +\infty,
\end{equation}
where $\alpha_r$ denotes the simple root of $E_7$ associated to the rightmost Kronecker subquiver in figure \ref{last2quivers}, we end up with a decoupled
pair $(Q_0,\cw_0)$ obtained by eliminating the two rightmost nodes.
Group theory suggests the pair $(Q,\cw)$ to correspond to a would--be $SU(2)\times SO(10)$ gauge theory coupled to $\tfrac{1}{2}(\mathbf{2},\mathbf{10})$. This is confirmed by a direct computation of the light BPS spectrum at weak coupling. Indeed, we get the same pattern as in \S\S.\ref{light26},\;\ref{light28}: the quantum numbers of the light vector--multiplets are given by the positive roots of $D_6$ which satisfy the constraint \eqref{ddim} (plus, again, one vector--multiplet corresponding to the $\mathfrak{su}(2)$ $W$--boson), while the other positive roots of $D_6$ give the quantum numbers of matter hypermultiplets. The vector--multiplets form the adjoint of $\mathfrak{so}(10)$, and the matter representations may be easily understood in terms of a fictitious Higgs breaking $SO(12)\rightarrow SO(10)$, confirming the expectation that they form the $\tfrac{1}{2}(\mathbf{2},\mathbf{10})$.

This result may seem odd, but there is no contradiction. The model is not UV complete, and there is no reason why it should be, given that the Higgs decoupled theory is not (in general) UV complete, even if we started from a sound QFT. The theory, however, exists as a low--energy effective theory of some UV completed microscopic theory. In \cite{cattoy} it was shown that such effective theories often have quivers, and that their light categories are nice, although the heavy sectors of the theory --- as the dyons --- are quite wild for a non--UV--complete theory \cite{cattoy}.

One can produce in the same way the pair $(Q,\cw)$ for the low energy effective theory with gauge group $SU(2)\times SO(2n)$ coupled to $\tfrac{1}{2}(\mathbf{2},\mathbf{2n})$ for any $n$.

\section{BPS spectrum at strong coupling: First results}
In this section we determine the spectrum of the simplest half-hypers models in a strongly coupled chamber by means of the mutation method of \cite{ACCERV2} (for a mathematical interpretation see \cite{kellerlast}). 
We limit ourselves to $SU(2)\times SO(2n)$ SYM coupled to $\tfrac{1}{2}(\mathbf{2},\mathbf{2n})$ leaving the more difficult models for future work.

We start by a brief review of the mutation method.

\subsection{Review of the mutation method \cite{ACCERV2}} The fact that a $\cn=2$ theory is quiver in the sense of \cite{CV11} entails, in particular, that there are always well defined cones of particles and anti--particles in the central charge plane, isomorphic via PCT. The choice of such a splitting is clearly a convention, physics being PCT invariant. Different choices, however, leads to different definitions of the quivers with potentials $(Q,\cw)$ we adopt to describe the BPS spectrum: all such quivers are S--equivalent, in the sense of section \S 2. The precise statement of such an equivalence is the core of the mutation method. Consider rotating the upper-half Z-plane $\ch \to e^{-i \theta} \ch$, for increasing values of $\theta \in [0,\pi)$. The leftmost central charge, say $Z_k$, will exit the particle cone and became part of the anti-particle one, simultaneously the charge $-Z_k$ will exit the anti-particle cone and enter the particle cone as the rightmost positive central charge. When this happens the quiver with superpotential we use to describe the theory undergoes a one dimensional Seiberg duality\footnote{\ In this case it is a left mutation. Instead, rotating the upper-half plane anti-clockwise, would produce a right mutation \cite{keller}: $\mu_L \circ \mu_R = \mu_R \circ \mu_L = \text{id}$. For simplicity, we are suppressing the index $L$.} at node $k$
\begin{equation}
(Q,\cw) \longrightarrow \mu_k(Q,\cw)
\end{equation}
the charges labeling the nodes of the quiver transform accordingly as
\begin{equation}
\mu_k \colon 
\begin{cases}
\mu_k(e)_k \equiv - e_k\\
 \mu_k(e)_j \equiv \begin{cases}
 e_j &\text{ if there are no arrows } k \to j \\ 
 e_j + m e_k & \text{ if there are } m \text{ arrows } k \to j
 \end{cases}
 \end{cases}
\end{equation}
If we are in a finite BPS chamber, by a $180^{\circ}$ rotation of $\ch$ we can reconstruct in this way a sequence of mutations
\begin{equation}
\boldsymbol{m} \equiv \mu_{k(s)} \circ  \mu_{k(s-1)} \circ \dots \circ \mu_{k(2)} \circ \mu_{k(1)},
\end{equation}
such that $i)$ it acts as the identity on the quiver with superpotential, and $ii)$ it acts as multiplication by $-1$ up to a permutation on the set of generators of the charge lattice $\Gamma = \oplus_{i=1}^n \mathbb{Z} e_i$:
\begin{equation}
\boldsymbol{m}(Q,\cw) = (Q,\cw) \quad\text{and}\quad\boldsymbol{m}(\{e_1,...,e_n\}) = \{- e_1,..., -e_n\}.
\end{equation}
Indeed, $e^{- i \pi}\ch$ has the same cone of particles of $\ch$ by PCT. By construction such a sequence of mutations encodes the (positive half) of the BPS spectrum ordered in decreasing central charge phase. The charges of the BPS hypermultiplets in this chamber are:
\begin{equation}\label{ppppqty}
\begin{aligned}
& e_{k(1)} ,\\
& \mu_{k(1)}(e)_{k(2)} , \\
& \mu_{k(2)} \circ \mu_{k(1)} (e)_{k(3)} ,\\
&\,\,\, \vdots \\
& \mu_{k(s-1)}\circ\dots\circ \mu_{k(2)}\circ \mu_{k(1)}(e)_{k(s)}.
\end{aligned}
\end{equation}

\subsection{$SU(2)\times SO(6)$ coupled to a half $(\mathbf{2},\mathbf{6})$.}
In order to describe a strongly coupled chamber for the model in consideration is better to use another quiver description of it, related to the one in eqn.\eqref{quiv26} by the sequence of mutations at nodes 1, 5, 3, 1, 3:
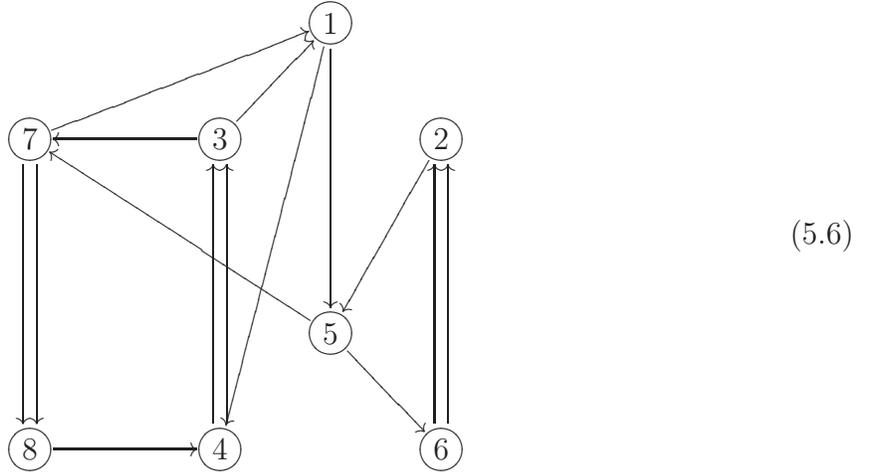
\begin{equation}\label{quiv26mu}
\begin{gathered}
\xymatrix{&& & *++[o][F-]{1}\ar[ddd]\ar[ddddl] &\\
*++[o][F-]{7}\ar@<0.5ex>@{->}[ddd]\ar@<-0.5ex>@{->}[ddd]\ar[urrr]&&*++[o][F-]{3} \ar[ll]\ar[ur]&&*++[o][F-]{2}\ar[ddl]\\
&&&&\\
&& &*++[o][F-]{5}\ar[dr]\ar[llluu]&\\
*++[o][F-]{8}\ar[rr]&&*++[o][F-]{4}\ar@<0.5ex>@{->}[uuu]\ar@<-0.5ex>@{->}[uuu] &&*++[o][F-]{6}\ar@<0.5ex>@{->}[uuu]\ar@<-0.5ex>@{->}[uuu] \\
}
\end{gathered}
\end{equation}
Notice that this is a very peculiar representative of the quiver mutation class of $SU(2)\times SO(6)$ coupled to a half $(\mathbf{2},\mathbf{6})$: The full subquiver on the complement of the nodes 7 and 8, in fact, is another representative of the $\mathcal{C}_{0,3}$ quiver mutation class. This is a further check of our `inverse Higgs' procedure. 
\smallskip

The sequence of mutations at nodes
\begin{equation}8,\ 5,\ 2,\ 3,\ 7,\ 2,\ 4,\ 1,\ 6,\ 3,\ 7,\ 1,\ 5,\ 2,\ 4,\ 5,\ 8,\ 1,\ 4,\ 5,\ 7,\end{equation}
 gives back the same quiver up to a permutation of nodes
\begin{equation}
 \pi=(1\,8\,6\,7\,4\,5\,2\,3)
\end{equation}
and acts on the generators $e_i$ of the charge lattice as\footnote{\ Although \textit{finding} the correct mutation sequence requires hard work, \emph{checking} it is very easy using the Keller mutation applet \cite{kellerappl}.} 
\begin{equation}\label{qqxxxxy}
 \boldsymbol{m}(e_{i})=-e_{\pi^{-1}(i)}.
\end{equation}
Then eqn.\eqref{ppppqty} gives a  finite BPS spectrum, consisting of 21 hypermultiplets with charges
\begin{equation}\label{spect26}
\begin{aligned}
&e_8, \quad e_5,\quad e_2,\quad e_3,\quad e_2+e_3+e_5+e_7,\quad e_3+e_5+e_7,\quad e_4+e_8,\\
&e_1+e_3+e_4+e_8,\quad e_1+e_3+e_4+e_5+e_6+e_8,\quad e_2+e_5+e_7,\\
&e_5+e_7,\quad e_5+e_6,\quad e_3+e_5+e_6+e_7,\quad e_5+e_6+e_7 \\
&e_1 +e_3,\quad  e_1,\quad  e_1+e_4,\quad e_3+e_7,\quad e_7,\quad e_4,\quad e_6.
\end{aligned}
\end{equation}
Such a spectrum, ordered in decreasing central charge arguments, satisfies opposite stability conditions compared with those implied by the weak coupling limit $g_2,g_4\rightarrow 0$.\smallskip

A different finite BPS chamber for this model,  with 27 hypers, will be described in the next subsection. 

\subsection{$SU(2)\times SO(8)$ coupled to a half $(\mathbf{2},\mathbf{8})$.}
By the sequence of mutations at nodes 1, 5, 3, 1, 3, the BPS quiver of this model is S-equivalent to the following one
\begin{equation}\label{quiv28mu}
\begin{gathered}
\xymatrix{&&&& & *++[o][F-]{1}\ar[ddd]\ar[ddddl] &\\
*++[o][F-]{10}\ar[rr]&&*++[o][F-]{7}\ar@<0.5ex>@{->}[ddd]\ar@<-0.5ex>@{->}[ddd]\ar[urrr]&&*++[o][F-]{3} \ar[ll]\ar[ur]&&*++[o][F-]{2}\ar[ddl]\\
&&&&&&\\
&&&& &*++[o][F-]{5}\ar[dr]\ar[llluu]&\\
*++[o][F-]{9}\ar@<0.5ex>@{->}[uuu]\ar@<-0.5ex>@{->}[uuu]&&*++[o][F-]{8}\ar[ll]\ar[rr]&&*++[o][F-]{4}\ar@<0.5ex>@{->}[uuu]\ar@<-0.5ex>@{->}[uuu] &&*++[o][F-]{6}\ar@<0.5ex>@{->}[uuu]\ar@<-0.5ex>@{->}[uuu] \\
}
\end{gathered}
\end{equation}

The sequence of mutations
\begin{multline}
10,\ 8,\ 3,\ 5,\ 2,\ 7,\ 2,\ 9,\ 4,\ 1,\ 10,\ 6,\ 7,\ 3,\ 10,\ 4,\ 2,\ 8,\ 3,\ 7,\ 2,\ 1,\ 3,\ 2,\ 5,\\
 4,\ 7,\ 8,\ 1,\ 10,\ 9,\ 1,\ 3,\ 5,\ 4,\ 1,\ 7,\ 4,\ 3,\ 8,\ 10,\ 2,\ 7,\ 3,\ 7,\ 1,\ 6,\ 10, 
\end{multline}
gives back the same quiver up to the permutation
\begin{equation}
 \pi=(1\,2\,5\,\,3\,8\,4\,10\,9\,6\, )(7)
\end{equation}
acting on the generators $e_i$ as
\begin{equation}
 \boldsymbol{m}(e_{i})=-e_{\pi^{-1}(i)}.
\end{equation}
This corresponds to a finite BPS chamber consisting of 48 hypers of charges
\begin{equation}\label{spect28}
\begin{aligned}
&e_{10}, \quad e_8, \quad  e_3, \quad  e_5, \quad  e_2, \quad  e_2 + e_3 + e_5 + e_7 + e_{10}, \quad e_3 + e_5 + e_7 + e_{10},\\
&e_8 + e_9, \quad e_4 + e_8, \quad e_1 + e_3 + e_4 + e_8 , \quad e_2 + e_3 + e_5 + e_7,\\
&e_1 + e_3 + e_4 + e_5 + e_6 + e_8, \quad e_3 + e_5 + e_7, \quad e_2 + e_5 + e_7 + e_{10},\\
&e_5 +e_7+e_{10}, \quad e_1 + e_3 , \quad  e_1 + e_2 +e_3 +e_5+e_7,\\
&e_1 + e_2+e_3+e_4+e_5+e_7+e_8+e_9, \quad e_1+e_3+e_5+e_7,\\ 
&e_1, \quad e_1 + e_3 + e_4+e_5+e_7+e_8+e_9, \quad 2 e_1 + 2 e_3 + e_4 + 2 e_5+ e_6 + e_7 + e_8 + e_9,\\
&e_1 + e_4+e_8+e_9, \quad 2 e_1 + e_3 + e_4 + e_5 + e_6 + e_8 + e_9, \quad e_3+e_7+e_{10},\\
&e_2+ e_3 + e_5 + 2 e_7 + e_{10}, \quad e_1 + e_8 + e_9, \quad e_3 + e_5 +2 e_7+e_{10}, \quad e_7 + e_{10},\\
&e_3 + e_7, \quad e_1+e_4+e_7+e_{10}, \quad e_1+e_4, \quad 2e_1+e_3+2 e_4 +e_5 +e_6+e_8+e_9,\\
&e_2+e_5+e_7, \quad e_5+e_7, \quad e_1+e_3+e_4+2e_5+e_6+e_7+e_8+e_9,\\
&2 e_1 + e_3 + e_4 + e_5+e_6,\quad e_1 + e_3 +e_4 + e_5 +e_6+e_8+e_9, \quad e_1 + e_3 + e_5 + e_6,\\
& e_1 + e_3 + e_5+e_6+e_7, \quad e_1 + e_5+e_6, \quad e_1 + e_4+e_5+e_6,\\
& e_5+e_6, \quad e_5+e_6+e_7, \quad e_7, \quad e_6, \quad e_9, \quad e_4.\\
\end{aligned}
\end{equation} 
From such a spectrum, we can deduce another mutation sequence for the model in the previous subsection, namely 
\begin{equation}8,\ 3,\ 5,\ 2,\ 4,\ 1,\ 7,\ 6,\ 2,\ 4,\ 3,\ 7,\ 2,\ 5,\ 8,\ 4,\ 3,\ 1,\ 8,\ 7,\ 5,\ 2,\ 1,\ 8,\ 1,\ 3,\ 5,\end{equation}
 that gives again the back the original quiver
up to the permutation
\begin{equation}
 \pi=(1\,2\,4\,5\,8\,6\,3\,7)
\end{equation}
while satisfying eqn.\eqref{qqxxxxy}.
The spectrum of such a chamber consists of 27 hypers of charges
\begin{equation}\label{spect26b}
\begin{aligned}
& e_8, \quad  e_3, \quad  e_5, \quad  e_2, \quad e_4 + e_8, \quad e_1 + e_3 + e_4 + e_8 , \quad e_2 + e_3 + e_5 + e_7,\\
&e_1 + e_3 + e_4 + e_5 + e_6 + e_8, \quad e_3 + e_5 + e_7, \quad e_1 + e_3 , \quad  e_1 + e_2 +e_3 +e_5+e_7,\\
& e_1+e_3+e_5+e_7, \quad e_1, \quad e_3 + e_7, \quad e_1+e_4, \quad e_2+e_5+e_7, \quad e_5+e_7,\\
& 2 e_1 + e_3 + e_4 + e_5+e_6, \quad e_1 + e_3 + e_5 + e_6, \quad e_1 + e_3 + e_5+e_6+e_7, \\
& e_1 + e_5+e_6, \quad e_1 + e_4+e_5+e_6, \quad e_5+e_6, \quad e_5+e_6+e_7, \quad e_7, \quad e_6, \quad e_4.\\
\end{aligned}
\end{equation} 
 Such a spectrum may be interpreted as a large--mass decoupling limit of the \eqref{spect28}.
Indeed, the quiver \eqref{quiv26mu} is a full subquiver of \eqref{quiv28mu}, and the spectrum of the chamber \eqref{spect26b} consists precisely of the representations of the quiver \eqref{quiv28mu} which are stable in the chamber \eqref{spect28} and have support on the  \eqref{quiv26mu} subquiver.

\bigskip

\bigskip

\section*{Acknowledgements}

We have greatly benefited of discussions with
Murad Alim,  William  Crawley--Boevey, Clay C\'ordova, Sam Espahbodi,
Ashwin Rastogi and Cumrun Vafa.  We thank them all.

\newpage
\appendix

\section{$\dim \C Q/(\partial\cw)<\infty$}\label{app:finite}

Here we show that the the Jacobian algebra 
$\C Q/(\partial\cw)$ for the $SU(2)\times SU(4)$ coupled to a half $(\mathbf{2},\mathbf{6})$ model (eqn.\eqref{quiv26}) is finite dimensional. This amounts to show that there is a finite set of words in the alphabet \begin{equation}\label{alphabet}\{H_1,\phi, V_1,H_3,\phi^\prime, \psi^\prime, A, B, \psi, H_2, V_2, V_3, h_1,h_2, h_3\}\equiv \{\alpha\}\end{equation} which, together with the 8 lazy paths, span the algebra. Only words which correspond to concatenated paths in the quiver \eqref{quiv26} are \emph{legal}. In the set of all legal words, we introduce an order $\prec$ corresponding to the lexicographic order of the words read in reverse (\textit{i.e.}\! from right to left) with respect to the `alphabetic order' in eqn.\eqref{alphabet}. Thus the words of the form $WH_1$, $W$ any word, will precede those of the form $W^\prime\phi$, which will precede the $W^{\prime\prime}V_1$ ones, \textit{ect.}. By the \textit{minimal word} for an element of $\C Q/(\partial\cw)$ we mean the one which is minimal with respect to $\prec$ in the equivalence class of legal words modulo the relations
\begin{equation}
 W\sim W^\prime\quad \Leftrightarrow\quad W^\prime=\lambda W+ \sum_\alpha \mu_\alpha W_\alpha^{(2)} (\partial_\alpha\cw)W_\alpha^{(1)},\quad \lambda,\in \C^\times, \mu_\alpha\in \C.
\end{equation}

Modulo equivalence, there are just 2 non--zero words of the form $WH_1$, namely $H_1$ and $H_2H_1$, since
\begin{align}
&H_3H_2H_1=\partial_{H_1}\cw\,H_1\sim 0\\
&\psi H_2H_1= \psi\big(\partial_{H_3}\cw-V_3\psi^\prime B\phi\big)\sim  (\psi V_3\psi^\prime)B\phi=-\partial_A\cw\, B\phi\sim 0.
\end{align}
Then there are just $11$ minimal non--zero words of the form $W\phi$. Indeed, in a legal word the second (from the right) letter should be $A$ or $B$ and the third $\psi^\prime$; the fourth is either $V_3$ or $h_2$. In the first case the next letter should be $H_3$ (since $\psi V_3\psi^\prime\sim 0$) and
\begin{equation}
 WH_3V_3\psi^\prime {A\choose B}\phi\sim W{H_3H_2V_2\psi^\prime B\choose V_1\phi^\prime}\phi\sim 0,
\end{equation}
while in the second case the next letter must be either  $h_1$ or $V_2$. The first possibility gives no minimal word, since $h_1h_2\psi^\prime\sim\phi^\prime \psi V_3\psi^\prime\sim 0$ while in the second case the following letters must be $H_2$ and $\psi$; now
\begin{equation}
W \psi H_2V_2h_2\psi^\prime {A\choose B}\phi \sim W {\phi H_3V_3\psi^\prime A\phi\choose \psi V_3 \psi^\prime A \phi}\sim W{\phi H_3(H_2V_2h_2\psi^\prime B+V_3h_3\phi^\prime)\phi\choose 0}\sim 0.
\end{equation}
We remain with the 11 words $\phi, A\phi, B\phi, \psi^\prime A \phi$, $\psi^\prime B\phi$, $V_3\psi^\prime A\phi$, $h_2\psi^\prime A\phi$, $h_2\psi^\prime B\phi$, $V_2h_2\psi^\prime A\phi$, $V_2h_2\psi^\prime B\phi$, $H_2V_2h_2\psi^\prime A\phi$.
Then the number of non--zero minimal words of the forms $WV_1$ and $WH_3$ is at most 14 (in facts less), since $W$ is necessarily of the form $W^\prime H_1$ or $W^{\prime\prime}\phi$.

Let us show that minimal words of the form $W\phi^\prime$ are finite in number. There are at most 14 minimal words of the form
$W^\prime V_1\phi^\prime$, then we have $h_3\phi^\prime$, $V_3h_3\phi^\prime$, and the words of the form
\begin{equation}W^{\prime\prime}\psi V_3h_3\phi^\prime= W^{\prime\prime}(\partial_{\phi^\prime}\cw-\phi V_1)\phi^\prime\sim W^{\prime\prime}\phi V_1 \phi^\prime
\end{equation}
which are at most 11. Then $\# \{W\phi^\prime\} \leq 19$.

Let us consider words of the form $W\psi^\prime$. We note that the words $W h_1h_2\psi^\prime\sim W \phi^\prime \psi V_3\psi^\prime\sim 0$. 
The non--trivial ones then have the forms $\psi^\prime$, $V_3\psi^\prime$, $W^\prime H_3V_3\psi^\prime$ (whose number is at most 14),
$h_2\psi^\prime$, $V_2h_2\psi^\prime$, $H_2V_2h_2\psi^\prime$, while those of the form $W^\prime \psi H_2 V_2 h_2\psi^\prime= W^\prime(\partial_B \cw-\phi H_3V_3\psi^\prime)$ are equivalent to a subeset of the $W^\prime H_3V_3\psi^\prime$ ones.
Then $\# \{W\psi^\prime\} \leq 28$.

This implies $\#\{ WA\}\leq 20$, $\# \{WB\}\leq 20$, $\# \{W\psi\}\leq 60$, $\#\{WH_2\}\leq 61$, $\# \{W V_2\}\leq 62$, $\# \{W V_3\}\leq 75$.
The words of the form $W^\prime h_ih_{i+1}$ are certainly not minimal, so the numbers of minimal words of the form $Wh_i$ are, respectively, at most 15, 63, 76. Thus we get the rough bound
\begin{equation}
 \dim \C Q/(\partial \cw)\leq 540.
\end{equation}

\section{Consistency with the perturbative Higgs effect}\label{lemma1}

For the convenience of the reader let us recall here the quiver of the $SU(2)\times SU(4)$ coupled to a half $(\mathbf{2},\mathbf{6})$ model:
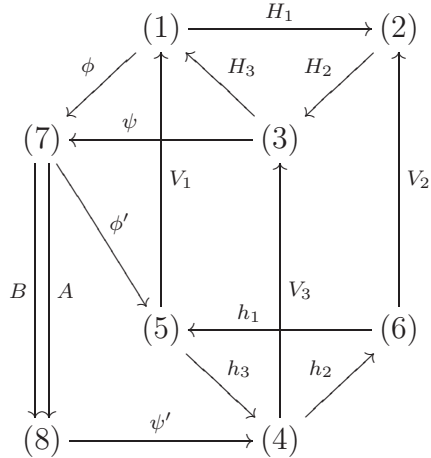
\begin{equation}
\begin{split}
\begin{gathered}
\xymatrix{&(1) \ar[ld]_{\phi} \ar[rr]^{H_1} && (2)\ar[dl]_{H_2}\\
(7) \ar@<0.5ex>@{->}[ddd]^{A}\ar@<-0.5ex>@{->}[ddd]_{B} \ar[ddr]^{\phi^{\prime}}& & (3)\ar[lu]_{H_3} \ar[ll]_{\psi \qquad} &\\
& &  &\\
 &(5)\ar[uuu]_{V_1}\ar[dr]^{h_3} & & (6)\ar[ll]_{h_1\qquad}\ar[uuu]_{V_2}\\
(8) \ar[rr]^{\psi^{\prime}}& & (4)\ar[ur]^{h_2}\ar[uuu]_{V_3} &\\
}
\end{gathered}&\qquad
\begin{aligned}
&\cw = H_1 H_3 H_2 + h_3 h_1 h_2 +\\
&+ A \psi V_3 \psi^{\prime} + B \psi H_2 V_2 h_2 \psi^{\prime} +\\
& + \phi V_1 \phi^{\prime} + \psi V_3 h_3 \phi^{\prime} +\\
& + \phi H_3 V_3 \psi^{\prime} B,
\end{aligned}
\end{split}
\end{equation}

{\bf Lemma.} If $X \in \mathscr{L}(Q,\cw)$ then
\begin{equation}
\begin{aligned}
&(1) \quad X \big|_{\mathbf{Kr}} \in \ct_{\mathbf{Kr}}\\
& (2) \quad X \big|_{\hat{A}(3,1)_i} \in \ct_{\hat{A}(3,1)} \quad \forall \, \, i=1,2,3
\end{aligned}
\end{equation}

\medskip

{\sc Proof.} The proof is by negation: We are going to prove that if the restriction of $X$ to any of the euclidean subdiagrams $\mathbf{Kr}$ or $\hat{A}(3,1)_i$ is not a direct sum of regulars, then $X \notin \mathscr{L}(Q,\cw)$. If  the restriction of $X$ is not a direct sum of regulars, then it must have an injective summand. We know from \cite{cattoy} that, for the Kronecker subquiver, this is equivalent to the existence of $\ell$ vectors $v_1, \dots, v_{\ell}$ such that
\begin{equation}
\begin{aligned}
& A \, v_1 = 0\\
& A \, v_{a+1} = B \, v_a \qquad a = 1, \dots, \ell -1\\
& 0 = B \, v_{\ell}
\end{aligned}
\end{equation}
As, far as the $\hat{A}(3,1)_i$ subquivers, the same is true replacing $A$ with $V_i$ and $B$ with the appropriate composition of arrows $H V h$. We are going to proceed by induction on $\ell$ for $\mathbf{Kr}$.

\medskip

The case $\ell =1$ is trivial: If $\exists$ $v \in X_7$ such that $A \, v = 0 = B \, v$, then, by $\partial_{\psi} \cw = 0$, $\partial_{h_1} \cw =0$, and $\partial_{\phi} \cw = 0$ we have that
\begin{equation}
V_3 h_3 \phi^{\prime}\, v = 0,\quad  h_2 h_3 = 0, \quad V_1\,\phi^{\prime} \, v = 0.
\end{equation}
So, there is always a subrep $Y$ with
\begin{equation}
Y_7 = \C v, \quad Y_5 = \C \phi^{\prime} \, v, \quad Y_4 = \C h_3 \phi^{\prime} \,  v, \quad Y_i = 0 \text{ else} 
\end{equation}
such that $m(Y) > 0$ and $X \notin \mathscr{L}(Q,\cw)$.

\medskip

Now, assume that this is true for $\ell -1$ vectors of $X_7$. We are going to prove that
\begin{equation}
\begin{aligned}
(i) \quad & \exists \, \ell \text{ such vectors for } X\big|_{\mathbf{Kr}} \Longrightarrow  \exists \, \ell \text{ such vectors for } X\big|_{\hat{A}(3,1)_3} \, ;\\
(ii) \quad & \exists \, M \text{ such vectors for } X\big|_{\hat{A}(3,1)_3} \Longrightarrow \exists \, M-1 \text{ such vectors for } X\big|_{\mathbf{Kr}}\, ;\\
(iii) \quad & \exists \, N \text{ such vectors for } X\big|_{\hat{A}(3,1)_1} \Longrightarrow \exists \, N-1 \text{ such vectors for } X\big|_{\mathbf{Kr}}\, .\\
\end{aligned}
\end{equation}
These implications entail that $X$ restricted to $\mathbf{Kr}$, $\hat{A}(3,1)_3$, and $\hat{A}(3,1)_1$ never has injective summands by the inductive hypothesis. \underline{Proof of $(iii)$}: In this case we have a family of $N$ vectors $v_a$ in $X_5$ such that
\begin{equation}
\begin{aligned}
& H_3 V_3 h_3 \, v_1 = 0\\
& H_3 V_3 h_3 \, v_{a+1} = V_1 \, v_a \qquad a = 1, \dots, N -1\\
& 0 = V_1 \, v_N.
\end{aligned}
\end{equation}
The set of $N-1$ vectors in $X_7$ is given by
\begin{equation}
z_a \equiv \psi V_3 h_3 v_a \qquad a = 1, \dots, N-1.
\end{equation}
If all $z_a = 0$, then, by $\partial_{\phi^{\prime}}\cw=0$, we have that $\phi V_1 v_a = 0$ for all $a$, therefore, since $\partial_{h_1} \cw = h_2 h_3 = 0$, such preinjective summand of $\hat{A}(3,1)_1$ would be a subrepresentation of $X$ with positive magnetic charge, so that $X \notin \mathscr{L}(Q,\cw)$. Assume therefore that $z_a \neq 0$ at least for some $a$. Composing to \eqref{mir2} $h_3$ on the right we obtain, since $h_2 h_3 = 0$,
\begin{equation}
A \psi V_3 h_3 + B \phi H_3 V_3 h_3 = 0
\end{equation}
Consider $B \phi H_3 V_3 h_3 \, v_a = B \phi V_1 \, v_{a-1} = - B \psi V_3 h_3 \, v_{a-1}$ for $a=2,\dots,N$, by $\partial_{\phi^{\prime}} \cw = \phi V_1 + \psi V_3 h_3 = 0$. This gives:
\begin{equation}
A z_1 = 0 \quad\text{and}\quad A z_a = - B z_{a-1} \quad a = 1,\dots, N-1.
\end{equation}
Then, since, again by $\partial_{\phi^{\prime}} \cw=0$,  $A \psi V_3 h_3 \, v_N = - A \phi V_1 v_N = 0$, we have that $B z_{N-1} = 0$. This concludes the proof of $(iii)$. \underline{Proof of $(i)$}: Given the set of $\ell$ vectors $v_a$ in $X_7$, define the set of $\ell$ vectors $w_a$ in $X_4$ as
\begin{equation}
w_a \equiv \psi^{\prime} A v_a + h_3 \phi^{\prime} \, v_a, \quad a=1,...,\ell.
\end{equation}
If the $w_a$ are all zero, then we have that $\psi^{\prime} B v_{a-1} + h_3 \phi^{\prime} \, v_a=0$, and, by  $\partial_{\phi} \cw = V_1 \phi^{\prime} + H_3 V_3 \psi^{\prime} B = 0$, $V_1 \phi^{\prime} v_a = - H_3 V_3 h_3 \phi^{\prime} v_{a+1}$, since, for stability, $\phi^{\prime} v_a$ cannot be zero for all $a$, we are back to case $(iii)$. Thus, we may assume $w_a \neq 0$ at least for some $a$. We have that, since $A v_1 = 0$,
\begin{equation}
H_2 V_2 h_2 \, w_1 = H_2 V_2 h_2 \, h_3 \phi^{\prime} \, v_1 = 0,
\end{equation}
by $\partial_{h_3} \cw = h_2 h_3 = 0$. Moreover, $H_2 V_2 h_2 \, w_a = H_2 V_2 h_2 \, \psi^{\prime} A v_a = H_2 V_2 h_2 \, \psi^{\prime} B v_{a-1}.$ Now, by the relation
\begin{equation}\label{mir1}
\partial_{\psi} \cw = V_3 \psi^{\prime} A + V_3 h_3 \phi^{\prime} + H_2 V_2 h_2 \psi^{\prime} B = 0,
\end{equation}
we have that $H_2 V_2 h_2 \, \psi^{\prime} B v_{a-1} = - V_3 w_{a-1}$, so
\begin{equation}
H_2 V_2 h_2 \, w_a = - V_3 w_{a-1}.
\end{equation}
Last, by \eqref{mir1}, 
\begin{equation}
V_3 \, w_{\ell} = - H_2 V_2 h_2 \psi^{\prime} B v_{\ell} = 0
\end{equation}
since $B v_{\ell} =0$. This proves $(i)$. \underline{Proof of $(ii)$}: Assume we have $M$ vectors $v_a$ in $X_4$ such that
\begin{equation}
\begin{aligned}
& H_2 V_2 h_2 \, v_1 = 0\\
& H_2 V_2 h_2 \, v_{a+1} = V_3 \, v_a \qquad a = 1, \dots, M -1\\
& 0 = V_3 \, v_M.
\end{aligned}
\end{equation}
The set of $M-1$ vectors in $X_7$ is given by
\begin{equation}
z_a \equiv \psi V_3 v_a \qquad a = 1, \dots, M-1.
\end{equation}
If $z_a = 0$ for all $a$, as in case $(iii)$, we have that such an injective summand is a subrepresentation of the whole $X$, and so, it is destabilizing (indeed, $H_3 H_2 = 0$ and $h_1 h_2 = - \phi^{\prime} \psi V_3$) . Else, consider the relation
\begin{equation}\label{mir2}
\partial_{\psi^{\prime}} \cw = A \psi V_3 + B \psi H_2 V_2 h_2 + B \phi H_3 V_3 = 0
\end{equation}
by $\partial_{H_1} \cw = H_3 H_2 = 0$, we have that the last term is such that $B \phi H_3 V_3 \, v_a = B \phi H_3 \, H_2 V_2 h_2 \, v_{a+1} = 0$ $\forall$ $a = 1,.., M$. This gives
\begin{equation}
A z_1 = 0 \quad\text{and}\quad A z_a = B z_{a-1} \quad a=2,...,M-1.
\end{equation}
Now, $B z_{M-1} = B \psi V_3 v_{M-1} =B \psi V_3 H_2 V_2 h_2 \, v_M = 0$, by applying \eqref{mir2} to $v_M$. This proves $(ii)$.  So far with the inductive argument.

\medskip

Now, if $X\big|_{\hat{A}(3,1)_2}$ has an injective summand then for some $N$, there are vectors
\begin{equation}
\begin{aligned}
& H_1 V_1 h_1 \, v_1 = 0\\
& H_1 V_1 h_1 \, v_{a+1} = V_2 \, v_a \qquad a = 1, \dots, N -1\\
& 0 = V_2 \, v_N.
\end{aligned}
\end{equation}
We have that $\partial_{h_2} \cw = h_3 h_1 + \psi^{\prime} B \psi H_2 V_2 = 0$. $h_3 h_1 v_N = 0$ trivially, but $\psi^{\prime} B \psi H_2 V_2 \, v_a = \psi^{\prime} B \psi H_2 H_1 V_1 h_1 \, v_{a+1} = 0$, since $H_2 H_1 = - V_3 \psi^{\prime} B \phi$ and $\psi V_3 \psi^{\prime} = 0$ by the eom's wrt to $H_3$ and $A$ respectively. Therefore, $h_3 h_1 v_a = 0$ for all $a = 1, \dots, N$. Now, by $\partial_{\phi^{\prime}} \cw = \phi V_1 + \psi V_3 h_3 = 0$, $\phi V_1 h_1 v_a = 0$ for all $a$, and, last, by $\partial_{H_3} \cw =0$, we have that $H_2 H_1 V_1 h_1 v_a = 0$ for all $a$. This implies that if $X\big|_{\hat{A}(3,1)_2}$ has an injective summand, then this summand is a subrepresentation of $X$ itself, with positive magnetic charge, so that $X \notin \mathscr{L}(Q,\cw)$.

\begin{flushright}
$\square$
\end{flushright}

Now, consider the quiver \eqref{quiv28}:
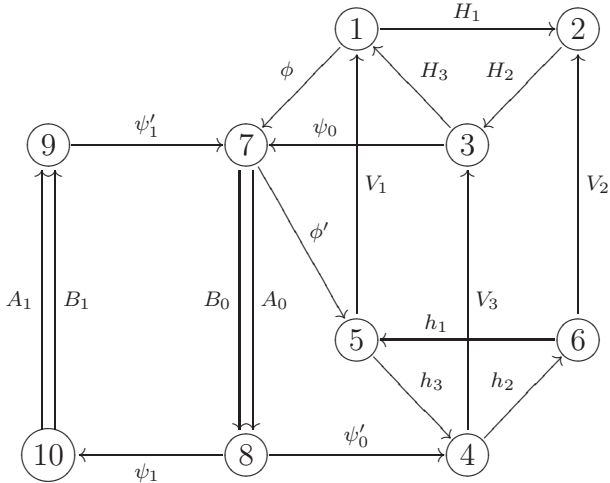
\begin{equation}
\begin{split}
\begin{gathered}
\xymatrix{&&&*++[o][F-]{1} \ar[ld]_{\phi} \ar[rr]^{H_1} && *++[o][F-]{2}\ar[dl]_{H_2}\\
*++[o][F-]{9}\ar[rr]^{\psi_1^{\prime}}&&*++[o][F-]{7} \ar@<0.5ex>@{->}[ddd]^{A_0}\ar@<-0.5ex>@{->}[ddd]_{B_0} \ar[ddr]^{\phi^{\prime}}& & *++[o][F-]{3}\ar[lu]_{H_3} \ar[ll]_{\psi_0 \qquad} &\\
&&& &  &\\
 &&&*++[o][F-]{5}\ar[uuu]_{V_1}\ar[dr]^{h_3} & & *++[o][F-]{6}\ar[ll]_{h_1\qquad}\ar[uuu]_{V_2}\\
*++[o][F-]{10} \ar@<0.5ex>@{->}[uuu]^{A_1}\ar@<-0.5ex>@{->}[uuu]_{B_1}&&*++[o][F-]{8}\ar[ll]^{\psi_1}  \ar[rr]^{\psi_0^{\prime}}& & *++[o][F-]{4}\ar[ur]^{h_2}\ar[uuu]_{V_3} &\\
}
\end{gathered}&\qquad
\begin{aligned}
&\cw = H_1 H_3 H_2 + h_3 h_1 h_2 +\\
&+ A_0 \psi_0 V_3 \psi_0^{\prime} + B_0 \psi_0 H_2 V_2 h_2 \psi_0^{\prime} +\\
& + \phi V_1 \phi^{\prime} + \psi_0 V_3 h_3 \phi^{\prime} +\\
& + \phi H_3 V_3 \psi_0^{\prime} B_0 + \\
&+ A_1 \psi_1 A_0 \psi_1^{\prime}  +  B_1 \psi_1 B_0 \psi_1^{\prime}\\
& = \cw_0 + A_1 \psi_1 A_0 \psi_1^{\prime}  +  B_1 \psi_1 B_0 \psi_1^{\prime},
\end{aligned}
\end{split}
\end{equation}

Reasoning in the same way as in the appendix of \cite{cattoy}, we can prove that if $X\big|_{\mathbf{Kr}_1}$ has an injective summand then, either $\psi^{\prime}$ is zero when restricted to such an injective summand, so that such a summand is a subrepresentation of the full $X$ with  $m_{K_1}(X)>0$ and $X\notin \mathscr{L}(Q,\cw)$, or we have the implication
\begin{equation}
\exists \, N \text{ such vectors for } X\big|_{\mathbf{Kr}_1} \Longrightarrow  \exists \, N-1 \text{ such vectors for } X\big|_{\mathbf{Kr}_0}
\end{equation}
Therefore, by the same inductive proof of the previous lemma, eqn.\eqref{secondfl} follows.

\section{We didn't forget anything}\label{pppqqw}

We want to show \emph{a priori} that  a representation $X$ of the quiver $Q^\prime$ \eqref{redSQM26}, bounded by the above relations,  with $X_0,X_3\neq0$ and is stable in the regime \eqref{chmb26} must satisfy
\begin{equation}
 \quad H_1=h_1=0.\label{dico}
\end{equation}
Indeed,
consider the following vector subspaces
\begin{equation}\label{subspaces}(0,\text{Im}(H_3\psi^\prime\psi),
\text{Im}(h_2\psi^\prime\psi),\text{Im}(\psi^\prime\psi))\subset (X_0,X_1,X_2,X_3)\end{equation}
Restricted to the respective subspaces as sources, one has $\psi|=\psi^\prime|=0$, while $H_3|$, $h_2|$ have images contained in the corresponding subspaces, and\footnote{\ For instance,
$$\text{Im }h_3|=\text{Im}(h_3H_3\psi^\prime\psi)=\text{Im}(h_3h_1h_2)=\text{Im}(\psi^\prime\psi H_2h_2)\subseteq \text{Im}(\psi^\prime\psi)$$
$$\text{Im }h_1|= \text{Im}(h_1h_2\psi^\prime\psi)=\text{Im}(H_3\psi^\prime\psi\psi^\prime\psi)=0.$$}\begin{equation}\text{Im }h_3|,\:\text{Im }H_2|\subseteq \text{Im}(\psi^\prime\psi),\qquad
\text{Im }H_1|=\text{Im }h_1|=0.\end{equation}
Therefore the subspaces \eqref{subspaces} form a subrepresentation $Y$ of $X$ which is destabilizing in our chamber, unless $Y$ is the zero object. Thus for a stable representation with $X_0,X_3$ non zero
$$\psi^\prime\psi=H_1H_3=h_3h_1=H_2H_1=h_1h_2=0.$$
Now consider the subspaces
$$ (\text{Im }\psi, X_1,X_2,X_3)\subseteq  (X_0, X_1,X_2,X_3).$$
From \eqref{chmb26} we see that they form a subrepresentation which is destabilizing for $X_0,X_3$ non zero, unless it coincides with $X$ itself. Thus $\psi$ is surjective and hence $\psi^\prime=0$.

Now, consider the following linear map
\begin{equation}
(X_0,X_1,X_2,X_3) \longmapsto (0,h_1 H_1 X_1, H_1h_1 X_2, 0).
\end{equation}
It is an element of End $X$ (whether $X$ is stable or not); indeed,
\begin{gather}
h_3h_1H_1=-\psi^\prime\psi H_2H_1=\psi^\prime\psi \psi^\prime\psi h_3=0\\
h_1H_1H_3=-h_1h_2\psi^\prime\psi=H_3\psi^\prime\psi \psi^\prime\psi=0\\
H_2H_1h_1=-\psi^\prime\psi h _3h_1=\psi^\prime\psi\psi^\prime\psi
H_2=0\\
H_1h_1h_2=-H_1H_3\psi^\prime\psi=h_2\psi^\prime\psi\psi^\prime\psi=0.
\end{gather}
A stable representation is, in particular, a brick; hence for a stable representation with  both $X_0$ and $X_3$ non zero,
\begin{equation}\label{rel26g}
h_1 H_1 = 0 = H_1 h_1.
\end{equation}
For such a stable representation then
\begin{equation}\begin{split}
&0 \to (0,0,\text{Im } H_1, 0) \to (X_0,X_1,X_2,X_3)\\
&0 \to (0,\text{Im } h_1, 0,0) \to (X_0,X_1,X_2,X_3)
\end{split}\end{equation}
are destabilizing subrepresentations unless they vanish. Then for a stable representation with $X_0,X_1\neq 0$, $h_1=H_1=0$.

\section{Proof of the identities used in \S.\,\ref{sec:univeraality}}\label{app:proofs}

For the superpotential in eqn.\eqref{geom620+}, the relations $\partial\cw^\prime=0$ are
\begin{gather}\label{IIdeide1}
  \psi^\prime A+(H_2h_2+h_3H_3)\psi^\prime=A\psi+\psi(H_2h_2+h_3H_3)=0\\
\widetilde{\psi}^\prime \widetilde{A}+(H_1h_1+h_2H_2)\widetilde{\psi}^\prime=\widetilde{A}\widetilde{\psi}+\widetilde{\psi}(H_1h_1+h_2H_2)=0\\
 \psi\psi^\prime=\widetilde{\psi}\widetilde{\psi}^\prime=0\label{IIdeide100}\\
H_3H_2+h_1\widetilde{\psi}^\prime\widetilde{\psi}=0\label{IIdeide106}\\
H_1H_3+h_2\psi^\prime\psi+\widetilde{\psi}^\prime\widetilde{\psi}h_2=0\\
H_2H_1+\psi^\prime\psi h_3=0\\
h_2h_3+\widetilde{\psi}^\prime\widetilde{\psi}H_1=0\\
h_3h_1+\psi^\prime\psi H_2+H_2\widetilde{\psi}^\prime\widetilde{\psi}=0\\
h_1h_2+H_3\psi^\prime\psi=0.
\end{gather}
from which it follows, in particular,
\begin{equation}\label{IIdeide10}
 (\psi^\prime\psi)^2=(\widetilde{\psi}^\prime\widetilde{\psi})^2=0.
\end{equation}

\subsection{Proof of eqn.\eqref{whatell}}\label{app:aa}

Eqn.\eqref{whatell} is equivalent to the equations
\begin{equation}
 \begin{gathered}
  \psi^\prime A+(H_2h_2+h_3H_3)\psi^\prime=A\psi+\psi(H_2h_2+h_3H_3)=0\\
\widetilde{\psi}^\prime \widetilde{A}+(H_1h_1+h_2H_2)\widetilde{\psi}^\prime=\widetilde{A}\widetilde{\psi}+\widetilde{\psi}(H_1h_1+h_2H_2)=0\\
h_i(h_{i+1}H_{i+1}+H_ih_i)=(H_{i-1}h_{i-1}+h_iH_i)h_i\\
H_i(h_iH_i+H_{i-1}h_{i-1})=(H_ih_i+h_{i+1}H_{i+1})H_i
 \end{gathered}
\end{equation}
where $i=1,2,3$ is defined mod 3. The first two lines are simply $\partial_\psi\cw=\partial_{\psi^\prime}\cw=\partial_{\widetilde{\psi}}\cw=\partial_{\widetilde{\psi}^\prime}\cw=0$.
The last two require $H_{i-1}h_{i-1}h_i=h_ih_{i+1}H_{i+1}$ and $H_iH_{i-1}h_{i-1}= h_{i+1}H_{i+1}H_i$.
Indeed
\begin{align*}
 &H_1h_1h_2= -H_1H_3\psi^\prime\psi=(h_2\psi^\prime\psi+\widetilde{\psi}^\prime\widetilde{\psi}h_2)\psi^\prime\psi=
\widetilde{\psi}^\prime\widetilde{\psi}h_2\psi^\prime\psi=-\widetilde{\psi}^\prime\widetilde{\psi}H_1H_3=h_2h_3H_3\\
&H_2h_2h_3=-H_2\widetilde{\psi}^\prime\widetilde{\psi}H_1=(h_3h_1+\psi^\prime\psi H_2)H_1=h_3h_1H_1-(\psi^\prime\psi)^2h_3=h_3h_1H_1\\
&H_3h_3h_1=-H_3(\psi^\prime\psi H_2+H_2\widetilde{\psi}^\prime\widetilde{\psi})=h_1h_2H_2+h_1(\widetilde{\psi}^\prime\widetilde{\psi})^2=h_1h_2H_2\\
&h_2H_2H_1= -h_2\psi^\prime\psi h_3=(H_1H_3+\widetilde{\psi}^\prime\widetilde{\psi}h_2)h_3=H_1H_3h_3\\
&h_3H_3H_2= -h_3h_1\widetilde{\psi}^\prime\widetilde{\psi}=\psi^\prime\psi H_2\widetilde{\psi}^\prime\widetilde{\psi}=
-\psi^\prime\psi h_3h_1= H_2H_1h_1\\
&h_1H_1H_3=-h_1(h_2\psi^\prime\psi+\widetilde{\psi}^\prime\widetilde{\psi}h_2)=H_3H_2h_2,
\end{align*}
where the relations \eqref{IIdeide106}--\eqref{IIdeide10} have been used repeatedly.

\subsection{Identities for bricks of \eqref{geom620}\eqref{geom620+}}

 If $X\in\mathsf{rep}(Q^\prime,\cw^\prime)$ is a brick, we have by \S.\,\ref{app:aa}
\begin{gather}\label{ideideK1}
A=\widetilde{A}=-\lambda\\
H_2h_2+h_3H_3=\lambda\label{ideideK2}\\
H_1h_1+h_2H_2=\lambda\\
H_3h_3+h_1H_1=\lambda\label{ideideK10}
\end{gather}
Any representation of $Q^\prime$ satisfying \eqref{ideideK1}--\eqref{ideideK10} automatically satisfies
$\partial_{\psi}\cw=\partial_{\psi^\prime}\cw=\partial_{\widetilde{\psi}}\cw=\partial_{\widetilde{\psi}^\prime}\cw=0$.
Then, at fixed $\lambda$, we may forget the loops in $Q^\prime$ and use the following quiver $\widetilde{Q}$
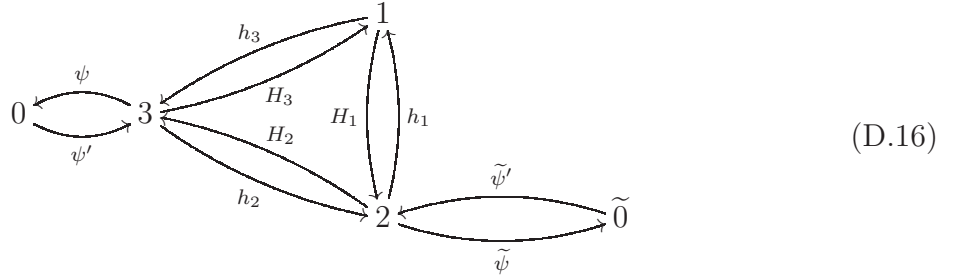
\begin{equation}\label{geomeeeAPP}
\begin{gathered}
\xymatrix@R=2.0pc@C=3.0pc{
&&&1 \ar@/_0.5pc/[dll]_{h_3} \ar@/_0.5pc/[dd]_{H_1}\\
  0 \ar@/_0.7pc/[r]_{\psi^{\prime}} & 3 \ar@/_0.7pc/[l]_{\psi} \ar@/_0.5pc/[urr]_{H_3} \ar@/_0.5pc/[drr]_{h_2}&&\\
&&&2 \ar@/_0.5pc/[ull]_{H_2} \ar@/_0.5pc/[uu]_{h_1}\ar@/_0.7pc/_{\widetilde{\psi}}[rr]&& \widetilde{0}\ar@/_0.7pc/_{\widetilde{\psi}^\prime}[ll]
}
\end{gathered}
\end{equation}
subjected to \eqref{ideideK2}--\eqref{ideideK10} as well as the relations \eqref{IIdeide100}--\eqref{IIdeide10}.
\medskip

\textbf{Lemma.} \textit{1) The relations $h_iH_i+H_{i-1}h_{i-1}=\lambda$ ($i$ defined mod 3) imply for $i=1,2,3$,
\begin{gather}
 h_i[H_ih_i(\lambda-H_ih_i)]=[H_{i-1}h_{i-1}(\lambda-H_{i-1}h_{i-1})]h_i\\
[H_ih_i(\lambda-H_ih_i)]H_i=H_i[H_{i-1}h_{i-1}(\lambda-H_{i-1}h_{i-1})].
\end{gather}
2) The relations \eqref{IIdeide100}--\eqref{IIdeide10} imply the following identities}
\begin{align}
\label{wwide1} H_3h_3(\lambda-H_3h_3)&=0\\
\label{wwide2}
H_2h_2(\lambda-H_2h_2)&=-\psi^\prime\psi H_2h_2\psi^\prime\psi\\
H_1h_1(\lambda-H_1h_1)&=-\widetilde{\psi}^\prime\widetilde{\psi}h_2H_2\widetilde{\psi}^\prime\widetilde{\psi}.
\label{wwide3}
\end{align} 

\textsc{Proof.}
1) Indeed $h_iH_i+H_{i-1}h_{i-1}=\lambda$ implies
\begin{equation}
 H_{i-1}h_{i-1}(\lambda-H_{i-1}h_{i-1})=h_iH_i(\lambda-h_iH_i)
\end{equation}
multiplying this identity by $h_i$ on the right and, respectively, $H_i$ on the left we get the desired identities.

2) For \eqref{wwide1}
\begin{multline}
 H_3h_3(\lambda-H_3h_3)=H_3h_3\,h_1H_1=H_3\big(-\psi^\prime\psi H_2-H_2\widetilde{\psi}^\prime\widetilde{\psi}\big)H_1=\\
=-
H_3\,\psi^\prime\psi(H_2H_1)-(H_3H_2)\widetilde{\psi}^\prime\widetilde{\psi}\,H_1
=H_3(\psi^\prime\psi)^2 h_3+h_1(\widetilde{\psi}^\prime\widetilde{\psi})^2H_1
=0,
\end{multline}
for \eqref{wwide2}
\begin{multline}
 H_2h_2(\lambda-H_2h_2)=H_2h_2h_3H_3=H_2\big(-\widetilde{\psi}^\prime\widetilde{\psi}H_1\big)H_3
=H_2\,\widetilde{\psi}^\prime\widetilde{\psi}\big(h_2\psi^\prime\psi+\widetilde{\psi}^\prime\widetilde{\psi}h_2\big)=\\
=
H_2\,\widetilde{\psi}^\prime\widetilde{\psi}\,h_2\psi^\prime\psi= -(h_3h_1+\psi^\prime\psi)h_2\psi^\prime\psi
=h_3H_3(\psi^\prime\psi)^2-\psi^\prime\psi h_2\psi^\prime\psi=-\psi^\prime\psi h_2\psi^\prime\psi,
\end{multline}
for \eqref{wwide3}
\begin{multline}
 H_1h_1(\lambda-H_1h_1)=H_1h_1h_2H_2=H_1\big(-H_3\psi^\prime\psi\big)H_2
=\big(h_2\psi^\prime\psi+\widetilde{\psi}^\prime\widetilde{\psi}h_2\big)\psi^\prime\psi H_2
=\widetilde{\psi}^\prime\widetilde{\psi}h_2\psi^\prime\psi H_2=\\ =-\widetilde{\psi}^\prime\widetilde{\psi}h_2(h_3h_1+H_2\widetilde{\psi}^\prime\widetilde{\psi})=
(\widetilde{\psi}^\prime\widetilde{\psi})^2H_1-\widetilde{\psi}^\prime\widetilde{\psi}h_2H_2\widetilde{\psi}^\prime\widetilde{\psi}=-\widetilde{\psi}^\prime\widetilde{\psi}h_2H_2\widetilde{\psi}^\prime\widetilde{\psi}.
\end{multline}

\subsubsection{Idempotents and isomorphisms}

The \textbf{Lemma} implies that the linear map
\begin{equation}
 X_0, X_{\widetilde{0}} \mapsto 0,\qquad X_i\mapsto h_iH_i(\lambda-h_iH_i)X_i\quad i=1,2,3 
\end{equation}
is a \emph{nilpotent} endomorphism of $X$, hence zero since $X$ is a brick.
Then, for $\lambda\neq 0$, we set
\begin{equation}
 P_i=h_iH_i/\lambda,\qquad Q_i=H_{i-1}h_{i-1}/h,
\end{equation}
 and the $P_i$, $Q_i$ are (semisimple) orthogonal complementary idempotenents
\begin{equation}
 P_i^2=P_i,\qquad Q_i^2=Q_i,\qquad P_iQ_i=0,\qquad P_i+Q_i=\mathrm{Id}_{X_i}.
\end{equation}
Therefore
\begin{equation}
 X_i=P_iX_i\oplus Q_iX_i,
\end{equation}
and from the two identities $H_iP_i=Q_{i+1}H_i$, $h_iQ_{i+1}=P_ih_i$, we infer that the maps
\begin{equation}
 \xymatrix{P_iX_i \ar@<0.6ex>[rr]^{H_iP_i}&& Q_{i+1}X_{i+1}\ar@<0.6ex>[ll]^{h_iQ_{i+1}/\lambda}}
\end{equation}
are inverse isomorphisms
\begin{equation}
 (\lambda^{-1} h_iQ_{i+1})(H_iP_i)=P_i= \mathrm{id}_{P_iX_i},\qquad
(H_iP_i)(\lambda^{-1} h_iQ_{i+1})=Q_{i+1}= \mathrm{id}_{Q_{i+1}X_{i+1}}.
\end{equation}
\medskip

\subsubsection{Preprojective relations}
It remains to prove eqns.\eqref{relRELxxy5}--\eqref{relRELxxy10}.
The first one is obvious
\begin{equation}
 \psi(P_3^2+Q_3^2)\psi^\prime=\psi\psi^\prime=0.
\end{equation}
The first eqn.\eqref{relRELxxy52} is equivalent to $(h_3H_3)\psi^\prime\psi(h_3H_3)=0$ and
\begin{multline}
h_3H_3\psi^\prime\psi h_3H_3=-h_3H_3H_2H_1H_3=-h_3(-h_1\widetilde{\psi}^\prime\widetilde{\psi})(-h_2\psi^\prime\psi-
\widetilde{\psi}^\prime\widetilde{\psi}h_2)=\\=
\psi^\prime\psi H_2\widetilde{\psi}^\prime\widetilde{\psi}h_2\psi^\prime\psi= -\psi^\prime\psi h_3h_1h_2\psi^\prime\psi=
\psi^\prime\psi h_3H_3(\psi^\prime\psi)^2=0.
\end{multline}
The second one is equivalent to $H_1h_1\widetilde{\psi}^\prime\widetilde{\psi}H_1h_1=0$ and
\begin{multline}
 H_1h_1\widetilde{\psi}^\prime\widetilde{\psi}H_1h_1=-H_1h_1h_2h_3h_1=-H_1(-H_3\psi^\prime\psi)(-\psi^\prime\psi H_2-H_2\widetilde{\psi}^\prime\widetilde{\psi})=-H_1H_3\psi^\prime\psi H_2\widetilde{\psi}^\prime\widetilde{\psi}=\\
=\widetilde{\psi}^\prime\widetilde{\psi}h_2\psi^\prime\psi H_2\widetilde{\psi}^\prime\widetilde{\psi}=-
\widetilde{\psi}^\prime\widetilde{\psi}h_2h_3h_1\widetilde{\psi}^\prime\widetilde{\psi}=(\widetilde{\psi}^\prime\widetilde{\psi})^2H_1h_1\widetilde{\psi}^\prime\widetilde{\psi}=0.
\end{multline}
Eqn.\eqref{relRELxxy53} follows from $\widetilde{\psi}\widetilde{\psi}^\prime=0$ together with
\begin{equation}
 \lambda^{-1}P_2h_2\cdot Q_3H_2= P_2^3=P_2=(1-Q_2^2).
\end{equation}
 Finally, eqn.\eqref{relRELxxy10} is equivalent to
\begin{equation}
 \lambda^3 Q_3\psi^\prime\psi Q_3^2+\lambda^2 Q_3 H_2\widetilde{\psi}^\prime\widetilde{\psi}P_2h_2=0.
\end{equation}
Explicitly, the \textsc{lhs} is 
\begin{multline}
 H_2h_2\psi^\prime\psi H_2h_2H_2h_2+H_2h_2H_2\widetilde{\psi}^\prime\widetilde{\psi}h_2H_2h_2= H_2h_2[\psi^\prime\psi H_2+H_2\widetilde{\psi}^\prime\widetilde{\psi}]h_2H_2h_2=- H_1h_2h_3h_1h_2H_2h_2=\\ =
-H_2(\widetilde{\psi}^\prime\widetilde{\psi}H_1)(H_3\psi^\prime\psi)H_2h_2=H_2\widetilde{\psi}^\prime\widetilde{\psi}(h_2\psi^\prime\psi+\widetilde{\psi}^\prime\widetilde{\psi}h_2)\psi^\prime\psi H_2h_2=0.
\end{multline}

\end{document}